\documentclass[preprint,aps,preprintnumbers,nofootinbib,superscriptaddress,showpacs]{revtex4}
\usepackage{amsmath,amssymb,amsbsy}
\usepackage{graphicx}
\usepackage{epsfig}

\def\D{{\Delta}}

\def\S{{\Sigma}}

\def\t{{\tau}}

\newcommand*{\CPT}{\raise0.4ex\hbox{$\chi$}PT}
\newcommand*{\chpt}{\raise0.4ex\hbox{$\chi$}PT}
\newcommand*{\schpt}{S\raise0.4ex\hbox{$\chi$}PT}

\def\c#1{{\mathcal #1}}

\def\eqref#1{{(\ref{#1})}}

\def\CP{{\cal P}}

\def\CL{{\cal L}}
\def\CO{{\cal O}}

\def\CU{{\cal U}}
\def\CM{{\cal M}}

\def\MeV{\mathop{\rm MeV}\nolimits}

\def\bar{\overline}

\def\tilde{\widetilde}

\def\Str{\textrm{Str}}

\def\bk{$B_K\;$}

\def\bea{\begin{eqnarray}}
\def\eea{\end{eqnarray}}

\begin{document}

\preprint{CU-TP-1153}

\preprint{FERMILAB-PUB-06-325-T}

\vphantom{}

\title{The Kaon $B$-parameter in \\Mixed Action Chiral Perturbation Theory}

\author{C.\ Aubin}
\email[]{caubin@phys.columbia.edu}
\affiliation{Department of Physics, Columbia University, New York, NY 10027}

\author{Jack Laiho}
\email[]{jlaiho@fnal.gov}
\affiliation{Theoretical Physics Department, Fermilab, Batavia, IL 60510}

\author{Ruth S. Van de Water}
\email[]{ruthv@fnal.gov}
\affiliation{Theoretical Physics Department, Fermilab, Batavia, IL 60510}

\date{\today}

\begin{abstract}
We calculate the kaon B-parameter, $B_K$, in chiral perturbation
theory for a partially quenched, mixed action theory with
Ginsparg-Wilson valence quarks and staggered sea quarks. We find
that the resulting expression is similar to that in the continuum,
and in fact has only two additional unknown parameters.  At
one-loop order, taste-symmetry violations in the staggered sea
sector only contribute to flavor-disconnected diagrams by
generating an $\CO(a^2)$ shift to the masses of taste-singlet
sea-sea mesons.  Lattice discretization errors also give rise to
an  analytic term which shifts the tree-level value of \bk by an
amount of $\CO(a^2)$.   This term, however, is not strictly due to
taste-breaking, and is therefore also present in the expression
for \bk for pure G-W lattice fermions.   We also present a
numerical study of the mixed $B_K$ expression in order to
demonstrate that both discretization errors and finite volume
effects are small and under control on the MILC improved staggered
lattices.
\end{abstract}

\maketitle

\section{Introduction}
\label{sec:Intro}

Lattice quantum chromodynamics allows nonperturbative calculations
of low-energy QCD quantities from first principles.  Until
recently, limited computing resources and the inability to include
quark loop effects have prevented lattice calculations from
achieving realistic results. In the past few years, however,
lattice simulations of both light meson and heavy-light meson
quantities with dynamical staggered quarks have shown excellent
numerical agreement with experimental results
\cite{Davies:2003ik}.  These have included both post-dictions,
such as the pion and kaon decay constants \cite{Aubin:2004fs}, and
predictions, as in the case of the $B_c$ meson mass
\cite{Allison:2004be}.  Such successes demonstrate that many of
the systematic uncertainties associated with lattice simulations
are under control, and therefore give confidence that lattice
simulations can reliably calculate quantities that cannot be
accessed experimentally. One of the simplest quantities of
phenomenological importance that can only be calculated using
lattice QCD is the kaon B-parameter, $B_K$.  Thus a precise
measurement of $B_K$ is an important goal for the lattice
community.\footnote{Promising calculations of \bk including
dynamical quark effects are currently in progress using both
improved staggered fermions \cite{Gamiz:2006sq,Kim:2005st} and
domain-wall fermions \cite{Cohen:2006xi}.}

\bk parameterizes the hadronic contribution to  $K^0-\bar{K}^0$
mixing;  it therefore plays a crucial role  in extracting
information about the CKM matrix using measurements of the neutral
kaon system.  In particular, the size of indirect CP violation in
the neutral kaon system, $\epsilon_K$, when combined with a
numerical value for $B_K$, places an important constraint on the
apex of the CKM unitarity triangle \cite{UTfit, CKMfitter}.
Because $\epsilon_K$ is well known experimentally
\cite{Eidelman:2004wy}, the dominant source of error in this
procedure is the uncertainty in the lattice determination of
$B_K$.  It is likely that new physics would give rise to
CP-violating phases in addition to that of the CKM matrix;  such
phases would manifest themselves as apparent inconsistencies among
different measurements of quantities which should be identical
within the standard CKM picture.  Thus a precise determination of
$B_K$ will help to constrain physics beyond the standard model.

Because lattice simulations with staggered fermions can at present
reach significantly lighter quark masses than those with other
standard discretizations  \cite{Bernard:2001av, Aubin:2004wf},
calculations of weak matrix elements using the available
2+1-flavor Asqtad staggered lattices appear promising.
Unfortunately, however, one pays a significant price for the
computational speed of staggered simulations:  each flavor of
staggered quark comes in four species, or ``tastes.''  In the
continuum limit, these species become degenerate and can be
removed by taking the fourth root of the fermion determinant.   In
practice, however, one must take the fourth root of the quark
determinant during the lattice simulation in order to remove the
extra tastes; thus it is an open theoretical question whether or
not one recovers QCD after taking the continuum limit of
fourth-rooted lattice simulations. Recently several papers have
appeared addressing the validity of the ``fourth-root
trick"~\cite{Bernard:2006zw,Creutz:2006ys,Bernard:2006vv,Bernard:2006ee,Shamir:2006nj}.
Although the issue has not yet been resolved, there are
indications that the fourth root does not introduce pathologies
when taking the continuum limit of the lattice theory. We
recommend a recent review by Sharpe, Ref.~\cite{SharpePlenary}, as
a clear summary of the current status of the fourth-root trick. In
this paper we assume the validity of the fourth-root trick. Even
working under this assumption, however, the additional tastes
introduce complications to staggered lattice simulations.  The
degeneracy among the four tastes is broken by the nonzero lattice
spacing, $a$, and results in discretization errors of $\CO(a^2)$.
Thus one must use functional forms calculated in staggered chiral
perturbation theory (\schpt), in which taste-violating effects are
explicit, to correctly extrapolate staggered lattice data
\cite{Lee:1999zx,Aubin:2003mg,Aubin:2003uc,Sharpe:2004is}.

Staggered chiral perturbation theory has been used to successfully
fit quantities such as $f_\pi$ and $f_D$, for which the \schpt\
expressions are simple and taste-symmetry breaking primarily
enters through additive corrections to the pion masses inside
loops \cite{Aubin:2003uc,Aubin:2005aq}.  In the case of $B_K$,
however, the additional tastes also make the matching procedure
between the lattice $\Delta S=2$ effective four-fermion operator
and the desired continuum operator more difficult.  The latticized
version of the continuum \bk operator mixes with all other lattice
operators that are in the same representation of the staggered
lattice symmetry group \cite{Verstegen:1985kt} -- including those
with different tastes than the valence mesons.  Current staggered calculations only
account for mixing with operators of the correct taste and only to
1-loop order in $\alpha_S$;  this results in a ~20\% uncertainty
in \bk \cite{Gamiz:2006sq}.  In order to achieve a precise
determination of $B_K$ with staggered fermions, one must either
perform the lattice-to-continuum matching nonperturbatively using
a method such as that of the Rome-Southampton group \cite{Martinelli:1994ty}
or one must include the effects of the extra operators in the
\schpt\ expression  for $B_K$ \cite{VandeWater:2005uq}.  Thus
far the large number of staggered operators has prevented matching
calculations beyond 1-loop order \cite{Lee:2003sk,Becher:2005ke},
although nonperturbative matching including all relevant staggered
operators is, in principle, possible.  The effects of this
truncated lattice-to-continuum operator matching can be included
in an extended version of \schpt, but the resulting expression
for \bk has many undetermined fit parameters, only a few of which are
already known from measurements of other quantities.  Therefore
use of this expression may prove just as difficult as implementing
fully nonperturbative matching.

The calculation of weak matrix elements such as \bk with
Ginsparg-Wilson (G-W) quarks \cite{Ginsparg:1981bj}, on the other hand,
is theoretically much cleaner than that with staggered quarks.
This is because in the massless limit, G-W quarks
possess an exact chiral symmetry on the lattice and do not occur
in multiple species.  Although in practice, lattice simulations
use approximate G-W fermions, the degree to which chiral symmetry
is broken in simulations can be controlled either by the length of
the fifth dimension in the case of domain-wall quarks
\cite{Kaplan:1992bt, Shamir:1993zy} or through the degree to which
the overlap operator is realized in the case of overlap quarks
\cite{Narayanan:1992wx, Narayanan:1993sk, Narayanan:1994gw}.
Consequently, while the $\Delta S=2$ lattice operator still mixes
with wrong-chirality operators,  there are significantly fewer
operators than in the staggered case, and nonperturbative
renormalization can be used in the determination of $B_K$.  As an
additional benefit, the approximate chiral symmetry also ensures
that the appropriate chiral perturbation theory expression for use
in the extrapolation of \bk lattice data is continuum-like at next-to-leading order (NLO).
Lattice simulations with G-W fermions, however, are 10 to 100
times more computationally expensive than those with staggered
fermions with comparable masses and lattice spacing, and thus are unfortunately not yet practical for
realizing light dynamical quark masses.

A computationally affordable compromise is therefore to calculate
correlation functions with Ginsparg-Wilson valence quarks on a background of
dynamical staggered gauge configurations.
This ``mixed action" approach combines the advantages of staggered and G-W fermions while not suffering from their major disadvantages, and
is well-suited to the calculation of the weak matrix element $B_K$.
By using staggered sea
quarks and G-W valence quarks one can better approach the
chiral regime in the sea sector while minimizing operator mixing
and allowing the use of nonperturbative renormalization.
Additionally, because the MILC staggered lattices with 2+1 flavors
of dynamical quarks are publicly available and offer a number of
quark masses and lattice spacings \cite{Bernard:2001av,
Aubin:2004wf}, one can perform unquenched three-flavor simulations
at the same cost as quenched G-W simulations.
Mixed action simulations have already been successfully used to study quantities of
interest to nuclear physics \cite{Bistrovic:2005fd, Beane:2005rj},
thus we expect that a similar method can be used to calculate $B_K$.

One might be concerned that the use of a mixed action could
introduce new theoretical complications into the lattice
determination of $B_K$.   Although the mixed staggered sea, G-W
valence lattice theory reduces to QCD in the continuum limit, at
nonzero lattice spacing, it is manifestly unphysical in that it
violates unitarity.  Thus, in order to extract physical QCD
quantities from mixed action simulations, it is essential that one
can correctly describe and remove contributions  due to unphysical
mixed action effects from quantities such as $B_K$ using the
appropriate lattice chiral perturbation theory.  It has not been
rigorously proven that the mixed action chiral perturbation theory
developed in \cite{Bar:2005tu} is the correct chiral effective
theory for mixed G-W, staggered simulations in which the
fourth-root of the quark determinant is taken in the sea sector.
Nevertheless, Ref.~\cite{Bernard:2006zw} showed that, given a few
plausible assumptions, staggered chiral perturbation theory is the
correct chiral effective theory for describing the
pseudo-Goldstone boson sector of rooted staggered simulations.  A
similar line of reasoning should hold for mixed action chiral
perturbation theory.  Assuming so, mixed action lattice
simulations can be used to correctly calculate quantities
involving pseudo-Goldstone bosons (such as $B_K$) in QCD.  More
complicated quantities, however, should be considered on a
case-by-case basis.

\bigskip

In this paper we calculate \bk in \chpt\ for a lattice theory with
G-W valence quarks and staggered sea quarks. We present results
for a ``1+1+1" theory in which $m_u \neq m_d \neq m_s$ in the sea
sector, for a ``2+1" theory in which $m_u = m_d \neq m_s$ in the
sea sector, and for a ``full QCD"-like expression in which we tune
the valence-valence meson masses equal to the taste-singlet
sea-sea meson masses. (We emphasize, however, that these
expressions only truly reduce to QCD when $m_\textrm{sea} =
m_\textrm{valence}$ and the lattice spacing $a\to 0$.) These
expressions will be necessary for the correct chiral and continuum
extrapolation of mixed-action \bk lattice data.  We find that the
expression for \bk in mixed action \chpt\ has only two more
parameters than in the continuum.  The first coefficient
multiplies an analytic term which shifts the tree-level value of
\bk by an amount of $\CO(a^2)$;  this term is also present in the
case of pure G-W lattice fermions.  The second new parameter
shifts the mass of the taste-singlet sea-sea meson (which only
appears inside loop diagrams) by an amount of $\CO(a^2)$.  This
mass-splitting has already been separately determined, however, in the MILC spectrum calculations
so we do not consider it to be an unknown fit parameter. Therefore, in practice, the chiral
and continuum extrapolation of mixed action \bk lattice data
should be no more complicated than that of domain-wall lattice
data. A numerical analysis using the taste-breaking parameters
measured on the MILC coarse lattices ($a \approx 0.125$ fm) shows
that the size of non-analytic discretization errors should be less
than a percent of the continuum value of \bk over the relevant
extrapolation range. In addition, we find that finite volume
effects in \bk are also small, and are of $\CO(1\%)$ for the
lightest pion mass on the MILC coarse ensemble.

This paper is organized as follows.  We review mixed action chiral
perturbation theory (MA\chpt) in Sec.~\ref{sec:MAChPT}.  In
Sec.~\ref{sec:BKNLO} we calculate \bk to next-to-leading order in
MA\chpt.  This is divided into four subsections:  we first present
the spurion analysis necessary to map the quark-level \bk operator
onto an operator in the chiral effective theory in
Sec.~\ref{sec:BKSpur}, calculate the 1-loop contributions to \bk
at NLO in Sec.~\ref{sec:BK1-Loop}, determine the analytic
contributions to \bk at NLO in Sec.~\ref{sec:Anal}, and finally
present the complete expressions for \bk at NLO in MA\chpt\ in
Sec.~\ref{sec:Results}.  Next, in Sec.~\ref{sec:Num}, we estimate
the numerical size of both taste-symmetry breaking contributions
and finite volume effects on the existing MILC ensembles using the
resulting mixed action \chpt\ formulae.  Finally, we conclude in
Sec.~\ref{sec:Conc}.

\section{Mixed Action Chiral Perturbation Theory }
\label{sec:MAChPT}

In this section we review the leading-order mixed action chiral
Lagrangian, first determined in Ref.~\cite{Bar:2005tu}, and
discuss some of its physical consequences  for the
pseudo-Goldstone boson sector.

We consider a partially quenched theory with $N_\textrm{val}$
Ginsparg-Wilson valence quarks and $N_\textrm{sea}$ staggered sea
quarks.  Each staggered sea quark comes in four tastes, and each
G-W valence quark has a corresponding bosonic ghost partner to
cancel its contribution to loop diagrams.  For example, in the
case $N_\textrm{val} = 2$ and $N_\textrm{sea} = 3$ , the quark
mass matrix is given by
\begin{equation}\label{eq:Masses}
    M = \text{diag}(\underbrace{m_u, m_u, m_u, m_u, m_d, m_d, m_d, m_d, m_s, m_s, m_s, m_s}_\textrm{sea},
            \underbrace{m_x, m_y}_\textrm{valence},\underbrace{m_x, m_y}_\textrm{ghost}),
\end{equation}
where we label the dynamical quarks by $u,d,$ and $s$ and the
valence quarks by $x$ and $y$.  Near the chiral and continuum limits, the mixed action theory has an
approximate $SU(4N_\textrm{sea} +N_\textrm{val}
|N_\textrm{val})_{L} \otimes SU(4N_\textrm{sea} +N_\textrm{val}
|N_\textrm{val})_{R}$ graded chiral symmetry.
In analogy with QCD, we assume
that chiral symmetry spontaneously breaks to its vector subgroup,
\begin{equation}
    SU(4N_\textrm{sea} +N_\textrm{val} |N_\textrm{val})_{L} \otimes SU(4N_\textrm{sea} +N_\textrm{val} |N_\textrm{val})_{R} \xrightarrow{SSB} SU(4N_\textrm{sea}+N_\textrm{val} |N_\textrm{val})_{V},
\end{equation}
and gives rise to $(4 N_\textrm{sea} + 2 N_\textrm{val})^2 - 1$
pseudo-Goldstone bosons (PGBs).  These PGBs can be packaged in an
$SU(4N_\textrm{sea}+N_\textrm{val} |N_\textrm{val})$ matrix:
\begin{eqnarray}\label{eq:Sigma}
    \S = {\rm exp} \left( \frac{2 i \Phi}{f} \right), \;\;\;\;\;\;\;
    \Phi = \left( \begin{array}{cccccccc}
 U & \pi^+ & K^+&  Q_{ux} &  Q_{uy} &  \cdots & \cdots \\
 \pi^- & D & K^0& Q_{dx} & Q_{dy} & \cdots & \cdots \\*
 K^- & \bar{K}^0 & S &  Q_{sx} & Q_{sy} & \cdots & \cdots \\*
 Q_{ux}^\dagger & Q_{dx}^\dagger & Q_{sx}^\dagger &  X & P^+ & R_{\tilde xx}^\dagger    & R_{\tilde yx}^\dagger \\*
 Q_{uy}^\dagger  & Q_{dy}^\dagger  & Q_{sy}^\dagger   &  P^- & Y &R_{\tilde xy}^\dagger & R_{\tilde yy}^\dagger  \\*
\cdots & \cdots  & \cdots  & R_{\tilde xx} & R_{\tilde xy}  &
\tilde{X} & \tilde{P}^+ \\* \cdots & \cdots  & \cdots  & R_{\tilde
yx}  & R_{\tilde yy}  & \tilde{P}^- & \tilde{Y} \\*
\end{array}\right),
\end{eqnarray}
where $f$ is normalized such that $f_\pi \approx 131 \MeV$.  The
upper-left block of $\Phi$ contains sea-sea PGBs, each of which
comes in sixteen tastes.  For example,
\begin{equation}\label{eq:UDef}
U = \sum_{b=1}^{16}U_{b}\frac{T_{b}}{2}\ ,
\end{equation}
where the Euclidean gamma matrices
\begin{equation}
T_b = \{ \xi_5, i\xi_{\mu}\xi_{5}, i\xi_{\mu}\xi_{\nu},
\xi_{\mu},\xi_I\}
\end{equation}
are the generators of the continuum $SU(4)$ taste symmetry
($\xi_I$ is the $4\times 4$ identity matrix). The fields in the
central block are the flavor-charged ($P^+$ and $P^-$) and
flavor-neutral ($X$ and $Y$) valence-valence PGBs, while those in
the lower-right block with tildes are the analogous PGBs composed
of only ghost quarks. Finally, the off-diagonal blocks contain
``mixed" PGBs: those labelled by $R$'s are composed of one valence
and one ghost quark, while those labelled by $Q$'s are composed of
one valence and one sea quark.  We do not show the mixed ghost-sea
PGBs explicitly; their locations are indicated by ellipses.

Under chiral symmetry transformations, $\Sigma$ transforms as
\begin{equation}\label{eq:Sigma_trans}
    \S \longrightarrow L\ \S\ R^\dagger , \quad \quad
        L,R \in SU(4N_\textrm{sea}+N_\textrm{val} |N_\textrm{val})_{L,R}.
\end{equation}
The standard mixed action chiral perturbation theory
power-counting scheme is
\begin{equation}
    p_\textrm{PGB}^2 / \Lambda_\chi^2 \sim m_q / \Lambda_\textrm{QCD} \sim a^2 \Lambda_\textrm{QCD}^2\,,
\label{eq:epsilons}
\end{equation}
so the lowest-order, $\CO(p^2_\textrm{PGB}, m_q, a^2)$, mixed
action chiral Lagrangian is
\begin{equation}
    \c{L} = \frac{f^2}{8} \Str \big( \partial_\mu \S\, \partial_\mu \S^\dagger \big)
        - \frac{\mu f^2}{4} \Str \big( \S M^\dagger + M\S^\dagger \big)
        + a^2 \left( \c{U}_S +\c{U}^\prime_S + \c{U}_V \right),
\end{equation}
where $\Str$ indicates a graded supertrace over both flavor and
taste indices and $\mu$ is an undetermined dimensionful parameter
of $\CO(\Lambda_\textrm{QCD})$.   The leading-order expression for
the mass-squared of a valence-valence PGB is identical to that of
the continuum because the chiral symmetry of the valence sector
prevents an additive shift due to lattice spacing effects:
\begin{equation}
    m_{xy}^2 = \mu(m_x +m_y).
\label{eq:m_tree}\end{equation}
$\CU_S$ and $\CU_S'$ comprise the well-known staggered potential
and come from taste-symmetry breaking in the sea quark sector
\cite{Aubin:2003mg}.  $\CU_S$ splits the tree-level masses of the
sea-sea PGBs into degenerate  groups:
\begin{equation}
    {m}_{f f',t}^2 = \mu(m_{f} +m_{f'}) +a^2 \Delta_t,
\label{eq:mseasea}\end{equation}
where $\Delta_t$ is different for each of the $SO(4)$-taste
irreps:   $P$, $V$, $A$, $T$, $I$.  In particular, $\Delta_P = 0$
because the taste-pseudoscalar sea-sea PGB is a true lattice
Goldstone boson.   The mixed action Lagrangian contains only one
new operator, and thus one new low-energy constant, as compared to
the staggered chiral Lagrangian:
 \begin{equation}\label{eq:MixPotential}
    \c{U}_V = -C_\textrm{Mix}\ \Str \big( \t_3 \S \t_3 \S^\dagger \big),
\end{equation}
where
\begin{equation}
    \t_3 = \c{P}_\textrm{sea} -\c{P}_\textrm{val} = \text{diag}(I_\textrm{sea} \otimes I_\textrm{taste}, -I_\textrm{val}, -I_\textrm{val}).
\end{equation}
This operator links the valence and sea sectors and generates a
shift in the mass-squared of a mixed valence-sea PGB:
\begin{equation}\label{eq:ssvsmasses}
    {m}_{f x}^2 = \mu(m_f+m_x) +a^2 \D_\textrm{Mix}, \;\;\;\;\;\;
    \D_\textrm{Mix} = \frac{16 C_\textrm{Mix}}{f^2}\,.
\end{equation}
Although the parameter $\Delta_\textrm{Mix}$ has not yet been
calculated in mixed action lattice simulations, it can, in
principle, be determined by calculating the mass of a mixed valence-sea
meson on the lattice.  As in any partially quenched theory, the
mixed action theory contains flavor-neutral quark-disconnected
hairpin propagators which have double pole contributions.  The
only flavor-neutral propagators that appear in the expression for
\bk are those with two valence quarks;  these have the following
form:
\begin{eqnarray}\label{eq:prop}
    G_{XY}(q) = \frac{\delta_{XY}}{q^2 + m^2_{X}} + D_{XY}(q),
\end{eqnarray}
where
\begin{eqnarray}
    D_{XY}(q) &= & -\frac{1}{3} \frac{1}{(q^2 + m^2_{X})(q^2 + m^2_{Y})}
        \frac{(q^2 + m^2_{U_I})(q^2 + m^2_{D_I})(q^2 + m^2_{S_I})}{(q^2 + m^2_{\pi^0_I})(q^2 + m^2_{\eta_I})}
\label{eq:dis_prop}\end{eqnarray}
is the disconnected (hairpin) contribution.  Note that the sea-sea
PGBs in the above expression are all taste singlets because the
valence quarks do not transform under the taste symmetry.

\section{\bk at NLO in Mixed Action Chiral Perturbation Theory}
\label{sec:BKNLO}

In this section we outline the calculation of \bk in mixed action
\chpt.  We divide it into four subsections. We first determine the
operators that contribute to $B_K$ in the mixed action chiral
effective theory using a spurion analysis in
Sec.~\ref{sec:BKSpur}.  In Sec.~\ref{sec:BK1-Loop} we outline the
1-loop calculation of $B_K$, and in Sec.~\ref{sec:Anal} we follow
this up with an enumeration of the corresponding analytic terms.
Finally, the complete NLO results are presented in
Sec.~\ref{sec:Results}.

\subsection{\bk Spurion Analysis}
\label{sec:BKSpur}

The spurion analysis for \bk in the mixed
action case is similar to that in the continuum  \cite{Bijnens:1984ec,VandeWater:2005uq};
we will point out differences when they occur.

In continuum QCD, \bk is defined as a ratio of matrix elements:
\begin{equation}
    B_K \equiv \frac{\CM_K}{\CM_\textrm{vac}}\ .
    \label{eq:BKdef}
\end{equation}
The numerator in the above expression measures the hadronic contribution to neutral
kaon mixing:
\begin{eqnarray}
    \CM_K & = & \langle\bar{K}^0 | \CO_K | K^0\rangle,\\
    \CO_K &= & [\bar{s} \gamma_\mu (1 - \gamma_5)d][\bar{s}
    \gamma_\mu (1 - \gamma_5)d],
\end{eqnarray}
where we have dropped color indices in  $\CO_K$ because both color
contractions give rise to the same operators in the chiral
effective theory.  $\CO_K$ is an electroweak operator which
transforms as a $(27_L,1_R)$ under the standard continuum chiral
symmetry group. The denominator in Eq.~(\ref{eq:BKdef}) is the
same matrix element as in the numerator evaluated in the vacuum saturation approximation:
\begin{equation}
    \CM_K^{\rm vac} = \frac{8}{3}
    \langle \bar{K}^0 |\bar{s} \gamma_\mu (1 - \gamma_5) d
    |0\rangle \langle 0|\bar{s}\gamma_\mu(1 - \gamma_5) d|
    K^0\rangle ,
    \label{eq:MVac}
\end{equation}
so that \bk is normalized to be of $\CO(1)$.

In the mixed action theory, we define \bk in an analogous manner, except that
both the external kaons and the operator $\CO_K$ must now be
composed of valence quarks:
\begin{equation}
    \CO_K^\textrm{lat} = [\bar{y} \gamma_\mu (1 - \gamma_5)x][\bar{y} \gamma_\mu (1 - \gamma_5)x].
\end{equation}
We can rewrite this operator as follows:
\begin{equation} \label{eq:OK2}
    \CO_K^\textrm{lat} =
    4[\bar{q}_L (\gamma_\mu \otimes P_{\bar{y}x}) q_L]
    [\bar{q}_L(\gamma_\mu \otimes P_{\bar{y}x})  q_L]\, ,
\end{equation}
where the subscript ``$L$" on the quark field indicates the
left-handed projection and the matrix $P_{\bar{y}x}$ is a $4N_{\rm
sea}+2N_{\rm val}$ matrix in flavor space that projects out the
$\bar{y}x$ component of each quark bilinear ($[P_{\bar{y}x}]_{ij}
= \delta_{i,14} \delta_{j,13}$ for $N_{\rm sea}=3$, $N_{\rm
val}=2$).  If we promote $P_{\bar{y}x}$ to a spurion field, $F_K$,
which can transform under the mixed action chiral symmetry group
$SU(4N_{\rm sea}+N_{\rm val}|N_{\rm val})_L\otimes SU(4N_{\rm
sea}+N_{\rm val}|N_{\rm val})_R$, then $F_K$ must transform as
\begin{equation}\label{eq:FKtrans}
    F_K \to L F_K L^ \dagger\, ,
\end{equation}
so that Eq.~\eqref{eq:OK2} remains invariant.  One cannot build
a chirally invariant operator out of $\Sigma$ and the spurion field
$F_K$ without derivatives, but one can build two such operators at
$\CO(p_\textrm{PGB}^2)$:
\begin{eqnarray}
    && \sum_\mu \Str[\Sigma\partial_\mu\Sigma^\dagger F_K] \Str[\Sigma\partial_\mu\Sigma^\dagger F_K],  \\
   &&  \sum_\mu \Str[\Sigma\partial_\mu\Sigma^\dagger F_K \Sigma\partial_\mu\Sigma^\dagger F_K] .
\end{eqnarray}
It turns out however, that these two operators are equivalent when
one demotes the spurion $F_K$ to the matrix $P_{\bar{y}x}$.  Thus
we are left with a single chiral operator:
\begin{equation}\label{eq:BKchiral}
    \CO_K^\chi = \sum_\mu
    \Str[\Sigma\partial_\mu\Sigma^\dagger P_{\bar{y}x}]
    \Str[\Sigma\partial_\mu\Sigma^\dagger P_{\bar{y}x}]\, .
\end{equation}
This operator is identical in form to the continuum \bk operator
\cite{Bijnens:1984ec}, but $\Sigma$ contains more fields and the
standard trace has been promoted to a supertrace.

Because this operator is of $\CO(p_\textrm{PGB}^2)$, operators of
$\CO(m_q)$ and $\CO(a^2)$, if present, could also potentially
contribute at the same order in \chpt.  Recall that the quark mass
matrix, when promoted to a spurion field, must transform as $M\to
LMR^\dagger$; thus we cannot form a chiral operator with a single
power of $M$ and two powers of $F_K$ that is invariant under the
chiral symmetry.  This is, of course, to be expected because the
chiral symmetry of the G-W valence sector and the $U(1)_A$
symmetry of the staggered sea sector are sufficient to prevent any new
operators involving the quark mass matrix at leading order in the
chiral expansion.   New operators of $\CO(a^2)$ that are not present
in the continuum can also potentially appear and contribute to
$B_K$.  As discussed in Ref.~\cite{VandeWater:2005uq}, such
operators arise in two distinct ways:  mixing with
higher-dimension operators and insertions of four-fermion
operators from the action.  We demonstrate that these operators do
not introduce taste-symmetry breaking, and therefore give rise to the same
kinds of analytic terms as in the pure G-W case.

Although the \bk lattice operator is dimension 6, at the level of
the Symanzik effective theory, it maps onto all continuum
effective operators of dimensions 6 and higher that respect the
same lattice symmetries.  Operators of dimension 7 and 8 are
explicitly suppressed relative to the dimension 6 \bk  Symanzik
effective operator by powers of $a$ and $a^2$, respectively, and
can therefore be mapped onto chiral operators that may contribute
to \bk at NLO. Because, however, the lattice symmetry group
includes taste transformations under which the valence quarks are
singlets, only dimension 7 and 8 operators composed of four
valence quarks can possibly respect the same lattice symmetries as
the \bk lattice operator. Moreover, the chiral symmetry of the G-W
valence quarks in the \bk operator prohibits strictly valence
dimension 7 four-fermion operators.  Thus we need only consider
dimension 8 Symanzik effective operators for the \bk operator
which contain four valence quarks.
Fortunately we need not enumerate all possible dimension 8
quark-level operators of $\CO(a^2p^2)$ and $\CO(a^2m^2)$ in order
to determine all possible chiral operators of $\CO(a^2
p^2_\textrm{PGB})$ and $\CO(a^2 m_q)$ onto which they map. Because
the \bk lattice operator has a $L$-$L$ chiral structure, and
chiral symmetry is respected by the valence sector of the lattice
theory, it can only mix with higher-dimension operators that also
have a $L$-$L$ structure.  Consequently, these
dimension 8 four-fermion
operators will generate the same spurions as the dimension 6 \bk
operator, and thus lead to the same chiral operator as in
Eq.~(\ref{eq:BKchiral}), but with an additional undetermined
coefficient of $\CO(a^2)$.
In general, this new coefficient just produces an unknown shift of
$\CO(a^2)$ to the original $\CO(1)$ coefficient of the
$\CO_K^{\chi}$.  Thus it will not lead to any new functional forms
in the expression for $B_K$ in MA$\chi$PT, only additional
contributions from the leading-order $B_K$ operator that of are
higher-order in the MA$\chi$PT power-counting. In particular, at
NLO in mixed action $\chi$PT, it will simply lead to an $\CO(a^2)$
correction to the tree-level value of $B_K$, which we can absorb
into an analytic term.
We emphasize that, although this contribution is proportional to
$a^2$, it is not due to taste-symmetry breaking in the staggered
sea sector.  Because it arises from strictly valence four-fermion
operators, it is also present in simulations with pure G-W lattice
fermions.

New operators that contribute to \bk at $\CO(a^2)$ can also be
produced by inserting a dimension 6 $\CO(a^2)$ operator from the
Symanzik action into the \bk four-fermion operator.  A method for
combining four-fermion operators at the chiral level was developed
in Ref.~\cite{Sharpe:2004is} for the purpose of determining the
NLO staggered chiral Lagrangian. This method was later used in
Ref.~\cite{VandeWater:2005uq} to enumerate the operators that
arise from insertions of the staggered action that contribute to
\bk with both staggered sea and valence quarks.
In the case of the mixed action theory, we must consider
insertions of four-fermion operators with only sea quarks,
four-fermion operators with only valence quarks, and four-fermion
operators with both sea and valence quarks.

Let us first consider insertions of operators with only sea
quarks, as all of the work has essentially been done in
Ref~\cite{VandeWater:2005uq}. Staggered four-fermion operators can
be made invariant under arbitrary chiral symmetry transformations
by introducing six pairs of spurion fields:
\begin{eqnarray}\label{eq:stag_spur}
    && F_L \otimes F_L \rightarrow L F_L L^{\dagger} \otimes L F_L L^{\dagger}, \qquad
    F_R \otimes F_R \rightarrow R F_R R^{\dagger} \otimes R F_R R^{\dagger}, \nonumber\\
    && F_L \otimes F_R \rightarrow L F_L L^{\dagger} \otimes R F_R R^{\dagger}, \qquad \tilde{F}_L \otimes \tilde{F}_L \rightarrow L \tilde{F}_L R^{\dagger} \otimes L \tilde{F}_L R^{\dagger}, \nonumber\\
    && \tilde{F}_R \otimes \tilde{F}_R \rightarrow R \tilde{F}_R L^{\dagger} \otimes R \tilde{F}_R L^{\dagger}, \qquad
    \tilde{F}_L \otimes \tilde{F}_R \rightarrow L \tilde{F}_L R^{\dagger} \otimes R \tilde{F}_R L^{\dagger},
\end{eqnarray}
where the two separate spurions in each pair correspond to the two
separate quark bilinears in each four-fermion operator. The
taste-breaking spurions  in Eq.~(\ref{eq:stag_spur}) can be
combined with the \bk spurion $F_K$, which transforms as in
Eq.~(\ref{eq:FKtrans}), to produce all of the generic chiral
structures in Table~\ref{tab:InsOps}.\footnote{Note this table is
simply the first column of Table III in
Ref.~\cite{VandeWater:2005uq}.} In order to turn these structures
into operators that contribute to $B_K$, one must ultimately
replace the spurion fields with constant values.  In the case of
the staggered theory, because the constituent staggered bilinears
may have nontrivial tastes, these spurions become taste matrices,
$\xi_i \otimes \xi_i$, which are diagonal in flavor space. In the
case of the mixed action theory, because only the staggered sea
quarks carry taste quantum numbers, these spurions become taste
matrices multiplied by projectors onto the sea sector, i.e. $\xi_i
\otimes \xi_i \rightarrow \xi_i\CP_\textrm{sea} \otimes
\xi_i\CP_\textrm{sea}$. For example, consider a particular
operator in the mixed action theory coming from the first spurion
combination in Table~\ref{tab:InsOps}:
\begin{eqnarray}
    \Str(F_R \Sigma F_K \Sigma^{\dagger} F_R \Sigma F_K \Sigma^{\dagger}) + p.c. \xrightarrow{\textrm{sea-sea f-f. op.}} \Str(\xi_5 \CP_\textrm{sea} \Sigma P_{\bar{y}x} \Sigma^{\dagger} \xi_5 \CP_\textrm{sea} \Sigma P_{\bar{y}x} \Sigma^{\dagger}) + p.c.
\end{eqnarray}
Because this operator is already of $\CO(a^2)$, it can only
contribute to $B_K$ at NLO through tree-level diagrams;  thus it
can only contain two pion fields, which are insufficient to
separate all of the projectors onto the sea sector from all of the
projectors on the valence sector.   In fact, it is easy to show
that none of the chiral operators arising from insertions of
staggered four-fermion operators actually contribute to \bk at NLO
(although they may at higher orders) because their contributions
vanish identically due to the fact that
$\CP_\textrm{sea}P_{\overline{y}x}=0$.  Thus we do not show all of
their expressions here.

\begin{table}\begin{tabular}{lcl}  \hline\hline

\multicolumn{1}{c}{\emph{Generic Chiral Structure}}  \\[0.5mm] \hline

    $\Str(F_K \Sigma F_R\Sigma^\dagger F_K \Sigma F_R'\Sigma^\dagger) + p.c.$   \\[0.5mm]
    $\Str(F_K \Sigma F_R \Sigma^\dagger)\Str(F_K \Sigma F_R \Sigma^\dagger) + p.c.$   \\[0.5mm]

    \hline

    $\Str(F_K \tilde{F_L} \Sigma^\dagger F_K \Sigma \tilde{F_R}) + p.c.$   \\[0.5mm]
    $\Str(F_K \tilde{F_L} \Sigma^\dagger) \Str(F_K \Sigma \tilde{F_R}) + p.c.$   \\[0.5mm]
    $\Str(F_K \tilde{F_L} \Sigma^\dagger F_K \tilde{F_L} \Sigma^\dagger) + p.c.$   \\[0.5mm]
    $\Str(F_K \tilde{F_L} \Sigma^\dagger) \Str(F_K \tilde{F_L} \Sigma^\dagger) + p.c.$ &   \\[0.5mm]

    \hline\hline

\end{tabular}\caption{Mesonic operators corresponding to insertions of four-fermion
operators from the Symanzik effective action;  these generic
structures apply to operators with only sea quarks, to those with
only valence quarks, and to mixed operators with both sea and
valence quarks.  $F_K$ comes from the \bk quark-level operator and
will ultimately be set equal to the projector $P_{\bar{y}x}$.  The
remaining spurions come from four-fermion operators, and must be
set equal to different matrices depending upon the four-fermion
operator under consideration; the specific details are discussed
in the text.  The notation ``p.c.'' indicates the parity-conjugate
of the previous operator.}\label{tab:InsOps}\end{table}

We next consider insertions of mixed four-fermion operators in
which one quark bilinear contains staggered sea quarks and the
other contains GW valence quarks. Such mixed operators can be made
invariant under arbitrary chiral symmetry transformations with the
following four pairs of spurion fields~\cite{Bar:2005tu}:
\begin{eqnarray}\label{eq:mix_spur}
    && F_L \otimes F_L \rightarrow L F_L L^{\dagger} \otimes L F_L L^{\dagger}, \qquad
    F_R \otimes F_R \rightarrow R F_R R^{\dagger} \otimes R F_R R^{\dagger}, \nonumber\\
    && F_L \otimes F_R \rightarrow L F_L L^{\dagger} \otimes R F_R R^{\dagger}, \qquad F_R \otimes F_L \rightarrow R F_R R^{\dagger} \otimes L F_L L^{\dagger},
\end{eqnarray}
where, in this case, we have distinguished between the first
spurion in the pair (which corresponds to the sea bilinear) and
the second spurion (which corresponds to the valence bilinear).
This is necessary because the sea spurion will ultimately be
replaced with a projector onto the sea sector, while the valence
spurion will become a projector on the valence sector.   By
comparing Eqs.~(\ref{eq:mix_spur}) and (\ref{eq:stag_spur}), one
can see that the mixed spurion fields are actually a subset of the
staggered spurion fields, although they must be replaced with
different matrices in order to produce operators that contribute
to $B_K$.  Consequently, the only possible combinations of the \bk
spurion with the mixed spurions are already given in the upper
panel of Table~\ref{tab:InsOps}.   Although the mixed case is
clearly similar to the staggered one, let us nevertheless consider
the example of the operator arising from the first generic
structure in Table~\ref{tab:InsOps}:
\begin{eqnarray}
    \Str(F_R \Sigma F_K \Sigma^{\dagger} F_R \Sigma F_K \Sigma^{\dagger}) + p.c. \xrightarrow{\textrm{sea-val f-f. op.}} \Str(\CP_\textrm{sea} \Sigma P_{\bar{y}x} \Sigma^{\dagger} \CP_\textrm{val} \Sigma P_{\bar{y}x} \Sigma^{\dagger}) + p.c.
\end{eqnarray}
This operator does not contribute to \bk at NLO for the same
reasons as the previous example, and neither does the other
operator arising from insertions of mixed valence-sea four-fermion
operators.

Finally, we consider insertions of purely valence four-fermion
operators.  These operators lead to the same spurion fields as the
mixed operators~\cite{Bar:2005tu}, and thus to the same chiral
forms in the upper panel of Table~\ref{tab:InsOps}.  The only
difference is that both spurion fields must be replaced with
projectors onto the valence sector.   Consequently, as in the case
of the previous example, the new chiral operators do not
contribute to \bk at NLO. In summary,  although many MA$\chi$PT
operators of $\CO(a^2)$ arise from insertions of four-fermion
operators in the Symanzik effective action, none of them
contribute to \bk at NLO for the mixed G-W, staggered lattice
theory.

The previous analysis holds for G-W valence quarks, which have
perfect chiral symmetry.  On the lattice however, G-W quarks are
often approximated as domain-wall quarks, which have a small
amount of residual chiral symmetry breaking due to the finite size
of the fifth dimension. This chiral symmetry breaking is
parameterized by the residual mass, $m_{\rm res}$, which is a
measure of how far the left- and right-handed components of the
quarks extend into the fifth dimension. These effects can be
readily added to the chiral theory, as seen in
Refs.~\cite{Blum:2001sr,Blum:2001xb}, by adding the following
mass-like term to the
chiral
Lagrangian:
\begin{equation}
    \Delta \CL^\textrm{DWF} = - \frac{\mu f^2}{4} \Str \big(\Sigma \Omega^{\dagger} + \Omega \Sigma^{\dagger} \big),
\end{equation}
where
$\Omega$ is a spurion which
transforms as the mass matrix transforms, and in the end we set
$\Omega=m_\textrm{res} \times I $.
This leads to the familiar
expression for the tree-level mass of a PGB composed of two domain-wall quarks:
\begin{equation}
    m^2_{xy} = \mu(m_x + m_y + 2m_{\rm res}) .
\end{equation}
Clearly this term will not contribute at leading order to $B_K$,
since the $\Omega$ spurion transforms in the same manner as the
mass spurion, and the mass term did not contribute at this order.
Consequently, one may simply shift the valence quark masses by
$m_\textrm{val} \to m_\textrm{val} + m_\textrm{res}$ in the
results of this paper to transform them into expressions that
apply to lattice simulations with domain-wall valence quarks and
staggered sea quarks.\footnote{In applying $\chi$PT expressions to lattice simulations with domain-wall valence quarks and staggered sea quarks, it is also important to remember that the domain-wall valence and staggered sea quark masses are renormalized differently.  Consequently, if one wishes to use the bare domain-wall lattice Dirac mass parameter and the bare AsqTad staggered lattice mass in mixed action $\chi$PT expressions, one must allow the parameter $\mu$ which relates the quark masses to the pion mass-squared to be different in the valence and sea sectors.}

Finally we must consider the fact that, while one constructs a
lattice operator to correspond to a continuum operator with a
particular spin, once on the lattice, this operator is allowed to
mix due to gluon exchange with operators that correspond to other
continuum spin structures.
This comes from the fact that the symmetry group on the lattice is
not the $SO(4)$ group of Euclidean rotations: it has been broken
down to the subgroup of hypercubic rotations. Lattice operator
mixing patterns can become especially complicated when one
introduces staggered quarks because now the desired lattice
operator can not only mix with other operators with incorrect
spins, but also those with incorrect tastes.  Fortunately, in the
mixed action theory that we consider here, the symmetry of the G-W
valence sector is sufficient to prevent mixing between the lattice
\bk operator and new operators with nontrivial taste structure.
Because the valence quarks in the \bk four-fermion operator do not
transform under taste symmetry, the \bk operator is clearly not in
the same lattice symmetry irrep as any taste-violating
four-fermion operators.  Furthermore, in the case of pure G-W
valence quarks, the \bk operator cannot mix with operators of the
wrong chirality. In realistic simulations with domain-wall valence
quarks, however, the desired lattice \bk operator with spin
structure $VV+AA$ mixes with
four other operators which do not have the same $VV+AA$ spin
structure ($TT$, $VV-AA$, $SS+PP$, and $SS-PP$).  This
contamination though is suppressed by two factors of the residual
mass \cite{Aoki:2005ga}, and the effect is small, so long as the
residual mass is small. Although the effect may be small, it could
be non-negligible, but then it can be removed nonperturbatively
using the standard method of
Rome-Southampton~\cite{Martinelli:1994ty}.

\bigskip

To summarize, the result of this spurion analysis is that the
leading order operator that contributes to \bk in the mixed action
case is
simply the continuum operator naively generalized to the mixed
action theory. This is simpler than the full staggered case
\cite{VandeWater:2005uq},
in which many new operators appeared at leading order due to
taste-symmetry breaking. Nevertheless, taste-symmetry breaking
will still enter the mixed action calculation of \bk through the
masses of PGBs in loop diagrams.
In addition, taste-breaking operators will generate operators of
next-to-leading order (NLO) in the chiral effective theory that
will contribute to analytic terms; we will discuss these in
Sec.~\ref{sec:Anal}.

\subsection{Contribution of \bk at 1-Loop}
\label{sec:BK1-Loop}

Recall from Eq~(\ref{eq:BKdef}) that the kaon B-parameter is
defined as the ratio ${\cal M}_K/{\cal M}^{\rm vac}_K$. At
tree-level, \bea  \left(\frac{{\cal M}_K}{{\cal M}^{\rm
vac}_K}\right)^{LO} \equiv B_0. \eea Because all higher-order
contributions to \bk are identically zero in the limit of massless
quarks, this expression defines the B-parameter in the chiral and
continuum limits.

\begin{figure}
\begin{tabular}{ccccc}
    \epsfxsize=1.4in \epsffile{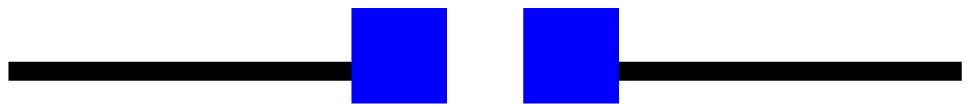} & $\;\;\;\;\;\;\;\;$ & \epsfxsize=1.4in \epsffile{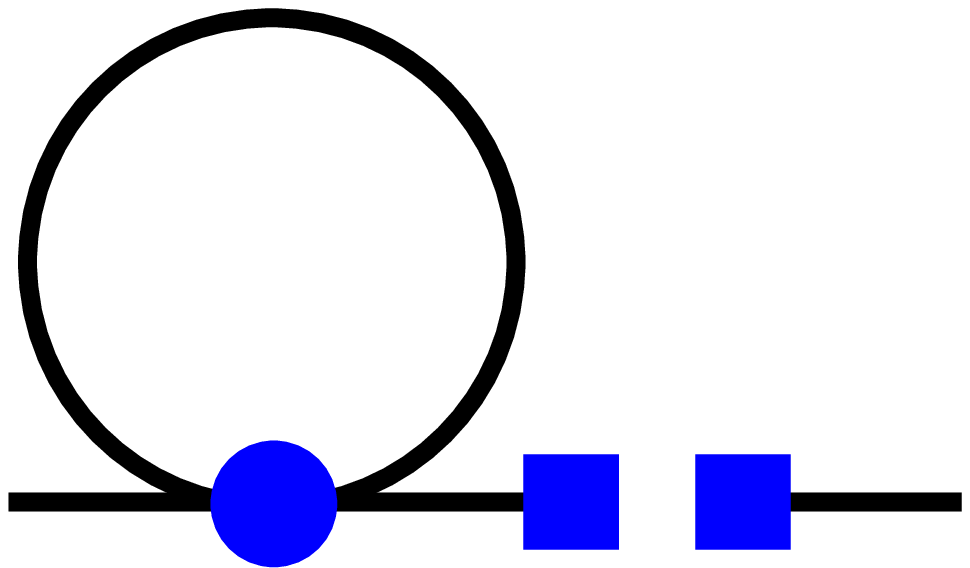} & $\;\;\;\;\;\;\;\;$ & \epsfxsize=1.4in \epsffile{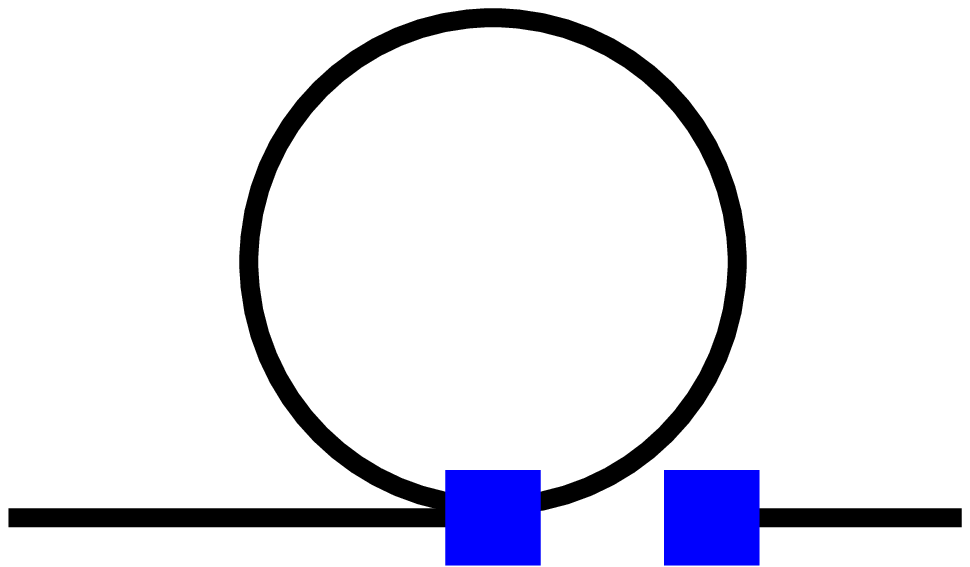} \\
    (a) & $\;\;\;\;\;\;$ & (b) & $\;\;\;\;\;\;$ & (c) \\
    \epsfxsize=1.4in \epsffile{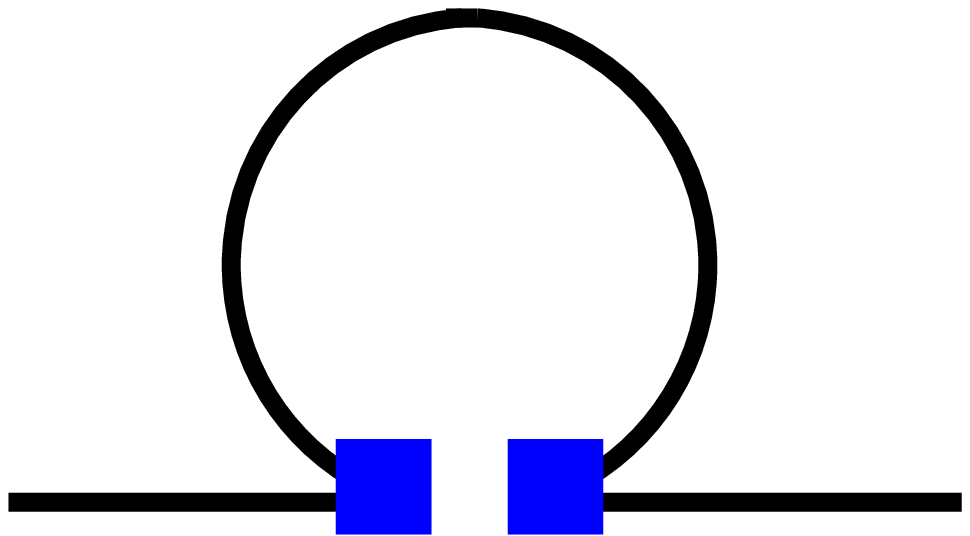} & $\;\;\;\;\;\;$ & \epsfxsize=1.4in \epsffile{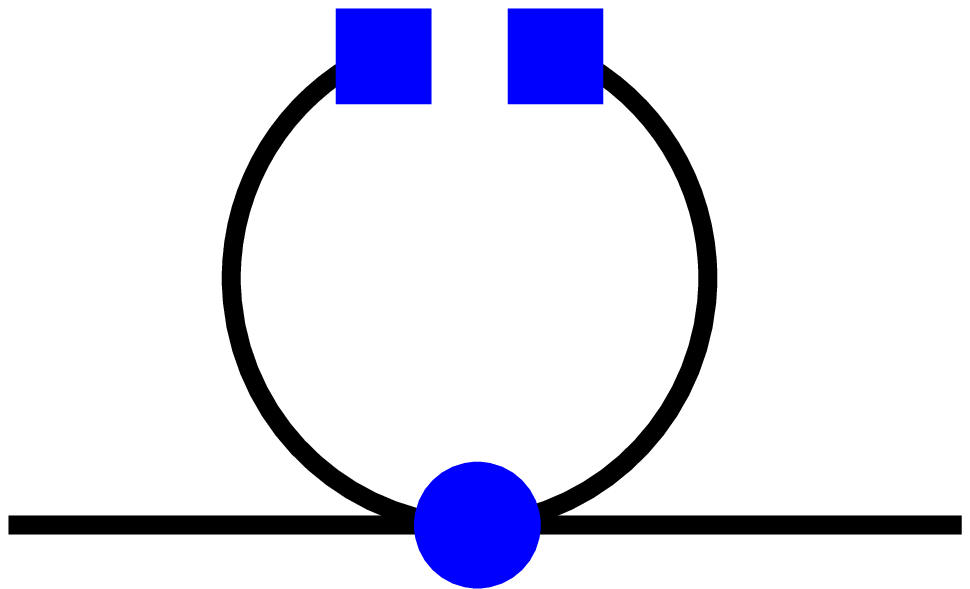} & $\;\;\;\;\;\;\;\;$ & \epsfxsize=1.4in \epsffile{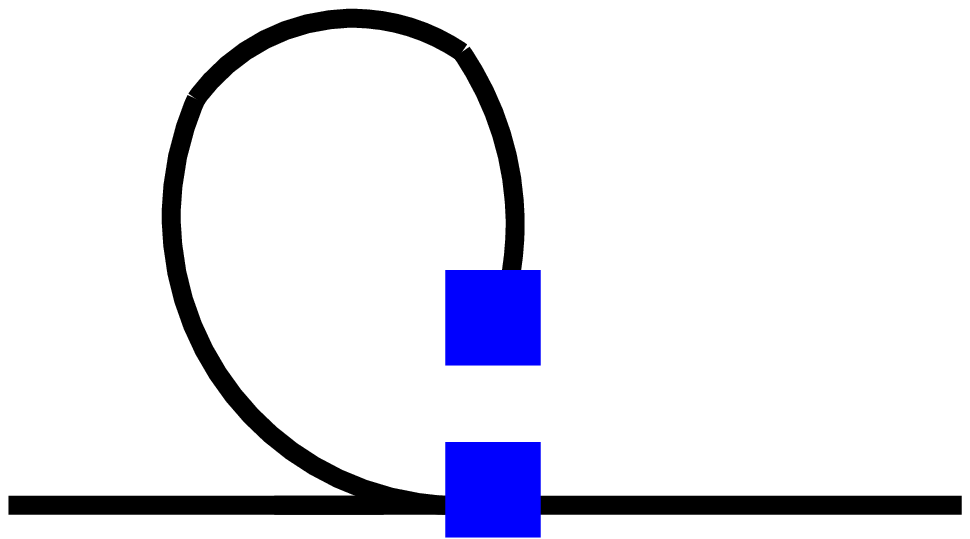} \\
    (d) & $\;\;\;\;\;\;$ & (e) & $\;\;\;\;\;\;$ & (f) \\
    \end{tabular}
    \caption{Tree-level and 1-loop contributions to $\CM_K$.  The circle represents a vertex from the LO staggered
     chiral Lagrangian.  Each square represents an insertion of one of the two left-handed currents in $\CO_K^\chi$ and ``changes" the quark flavor
    from $d \leftrightarrow s$. }\label{fig:BK1Loop}\end{figure}

At 1-loop, the $K^0-\overline{K}^0$ matrix element
receives contributions from the diagrams shown in
Fig.~\ref{fig:BK1Loop}, where we have specified the location of
each of the two left-handed currents in the chiral operator
$\CO_K^\chi$.\footnote{Note that this factorization is only
possible at leading order.  At higher orders, \bk receives
contributions from operators which are not simply products of
left-handed currents. } This factorization of the operator is
useful because the calculation of the kaon matrix element can then
be separated into two pieces, the 1-loop corrections to $f_K$ and
the 1-loop corrections to $B_K$ in which we are interested. In
terms of the contributions from the diagrams in
Figure~\ref{fig:BK1Loop}, the \bk matrix element can be written as
\bea\label{eq:MKdef} {\cal M}_K = \frac{8}{3}B_0 f^2
m^2_{xy}\{1+X[\textrm{Figs. 1(b)-(c)}] \} + X'[\textrm{Figs.
1(d)-(f)}]\}, \eea
where $X$ and $X'$ indicate the results of specific diagrams and
$m^2_{xy}$ is the 1-loop kaon mass squared.  At 1-loop, the form
of ${\cal M}^{\rm vac}_K$ is simple:
\bea {\cal M}^{\textrm{vac}}_K = \frac{8}{3}m^2_{xy}f^2_{xy}, \eea
where $m^2_{xy}$ and $f_{xy}$ are the 1-loop corrected values. It
is clear that diagrams (b) and (c) in Fig.~\ref{fig:BK1Loop}
factorize -- the left-half of each diagram is the 1-loop
correction to $f_K$, while the right half is just $f$ at
tree-level. Mathematically,
\bea X[\textrm{Figs. 1(b)-(c)}] = 2\frac{\delta f_{NLO}}{f}, \eea
where the factor of two comes from the fact that the
loop can appear on either leg.  Diagrams \ref{fig:BK1Loop}(b) and
\ref{fig:BK1Loop}(c) therefore turn the leading order $f^2$  into the 1-loop
$f^2_{xy}$ in Eq.~(\ref{eq:MKdef}):
\bea {\cal M}_K = \frac{8}{3}B_0 f^2_{xy}m^2_{xy} +
X'[\textrm{Figs. 1(d)-(f)}], \eea
such that $B_K$ at one loop only depends on diagrams
\ref{fig:BK1Loop}(d)-(f):
\bea B_K^{1-loop}=B_0 + \frac{3}{8} \frac{X'[\textrm{Figs.
1(d)-(f)}]}{f^2_{xy} m^2_{xy}}. \eea

\begin{figure}
\begin{tabular}{ccccc}
    \epsfxsize=1.4in \epsffile{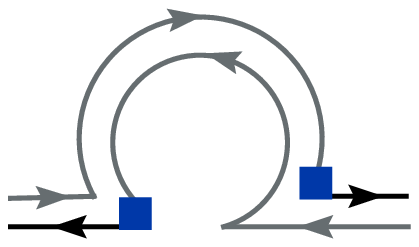} & $\;\;\;\;\;\;$ & \epsfxsize=1.4in \epsffile{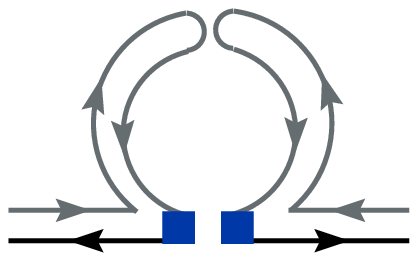} & $\;\;\;\;\;\;\;\;$ & \epsfxsize=1.4in \epsffile{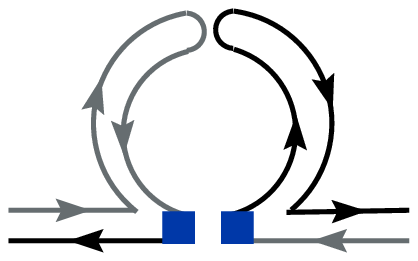} \\
    (a) & $\;\;\;\;\;\;$ & (b) & $\;\;\;\;\;\;$ & (c) \\
    \epsfxsize=1.4in \epsffile{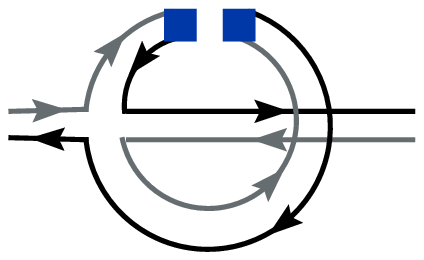} & $\;\;\;\;\;\;$ & \epsfxsize=1.4in \epsffile{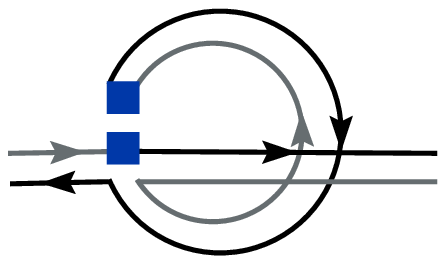} & $\;\;\;\;\;\;\;\;$ &  \\
    (d) & $\;\;\;\;\;\;$ & (e) & $\;\;\;\;\;\;$ &\\
    \end{tabular}
    \caption{Quark flow diagram contributions to $B_K$ at 1-loop.  One external
    meson is a $\bar{K}^0$ and the other is a $K^0$.  The two boxes represent an insertion
    of the $B_K$ operator.  Each box ``changes'' the valence quark flavor
    from $d \leftrightarrow s$.   Diagrams (a)--(c) contribute to Fig.~\ref{fig:BK1Loop}(d),
    diagram (d) contributes to Fig.~\ref{fig:BK1Loop}(e), and diagram (e) contributes
    to Fig.~\ref{fig:BK1Loop}(f).}\label{fig:BKQuark}\end{figure}

Figure~\ref{fig:BKQuark} shows the quark flow diagrams that
correspond to the meson diagrams in
Figs.~\ref{fig:BK1Loop}(d)-(f).  It is interesting to note that
the only place where sea quarks appear in these diagrams is in the
disconnected hairpin propagators of diagrams \ref{fig:BKQuark}(b)
and (c).  In particular, there are no contributions from mixed
mesons made of one sea and one valence quark, so the new parameter
in the mixed-action chiral Lagrangian, $\Delta_{\rm Mix}$, does
not appear in $B_K$ to 1-loop.\footnote{This cancellation of
chiral logarithms containing the parameter $\Delta_{\rm Mix}$ is
not unique to $B_K$. It also occurs in other mixed action chiral
perturbation theory expressions when they are written in terms of
1-loop PGB masses and decay constants (rather than bare
parameters), such as in the case of the $I=2$ $\pi\pi$ scattering
amplitude~\cite{Chen:2005ab}.}

We now proceed to calculate the 1-loop contributions to \bk from
Figure~\ref{fig:BKQuark}. The connected diagrams,
\ref{fig:BKQuark}(a), (d), and (e), combine to give the
result\footnote{We note that, although all of the integrals in
this section are divergent in four dimensions, one can choose a
suitable regulator to make them finite before their evaluation;
this regulator can then be removed after the results have been
renormalized.}
\bea \label{eq:Mconn}
{\cal M}_{conn} = \frac{B_0}{6 \pi^2}\int
\frac{d^4q}{(2\pi)^4}
\left[\frac{2m^4_{xy}}{q^2+m^2_{xy}}-\frac{m^2_X+m^2_{xy}}{q^2+m^2_X}-
\frac{m^2_Y+m^2_{xy}}{q^2+m^2_Y}\right]. \eea
The contribution from the disconnected diagrams,
\ref{fig:BKQuark}(b) and (c), is somewhat more tedious to evaluate
because of the double poles in the hairpin propagators:
\bea \label{eq:Mdisc}
{\cal M}_{disc}= \frac{2}{3}B_0\int
\frac{d^4q}{(2\pi)^4}(m^2_{xy}+q^2)\left\{D^I_{xx}(q)+D^I_{yy}(q)-2D^I_{xy}(q)
\right\}, \eea
where $D^I_{ij}$ is the taste-singlet disconnected propagator of
Eq~(\ref{eq:dis_prop}).  Making use of the identity
\cite{VandeWater:2005uq},
\bea D_{xx}+D_{yy}-2D_{xy} =
(m^2_X-m^2_Y)^2\frac{\partial}{\partial
m^2_X}\frac{\partial}{\partial m^2_Y}\{D_{xy}\}, \eea
we get the following result for the disconnected piece in the
``1+1+1" partially quenched theory:
\bea {\cal M}_{disc}^{PQ,1+1+1} =
\frac{B_0}{48\pi^2}(m^2_X-m^2_Y)^2\frac{\partial}{\partial
m^2_X}\frac{\partial}{\partial m^2_Y}\left\{\int
\frac{d^4q}{(2\pi)^4}\sum_{j}
\frac{(m^2_{xy}+m^2_j)}{(q^2+m^2_j)}R^{[4,3]}_j(\{M^{[4]}_{XY,I}\};\{\mu^{[3]}_I
\}) \right\}, \nonumber \\ && \eea
where $R^{[4,3]}_j$ is the residue arising from the double pole in
the disconnected propagator; $R^{[4,3]}_j$, $\{M^{[4]}_{XY,I}\}$, and $\{\mu^{[3]}_I\}$ are defined in Sec.~\ref{sec:Results}.

\bigskip

To get the full expression for \bk at NLO, one must combine the
1-loop contributions with analytic terms that arise from
tree-level matrix elements of higher-order operators:
\bea B^{PQ}_{K}= B_0 +\frac{3}{8}\left(\frac{{\cal M}_{conn}+{\cal
M}_{disc}}{f^2_{xy}m^2_{xy}}+\textrm{analytic terms} \right).
\eea
We discuss the analytic terms in the following subsection.

\subsection{Analytic contributions to \bk at NLO}
\label{sec:Anal}

Next-to-leading order analytic contributions to \bk come from
tree-level matrix elements of NLO operators.  In MA\chpt, such
terms can come from operators of the following order in the
power-counting scheme:
\bea
\CO(p_\textrm{PGB}^4),\ \CO(a^2 p_\textrm{PGB}^2),\ \CO(a^4),\ {\cal
O}(m^2_q),\ \CO(p_\textrm{PGB}^2 m_q),\ \CO(a^2 m_q). \eea
There are many such operators in the mixed action chiral Lagrangian,
however, since it is not necessary to separate them in fits to
numerical lattice data, we do not enumerate them all here. Instead
we use symmetry arguments to restrict the possible linearly
independent terms as in Ref.~\cite{VandeWater:2005uq}.

First we discuss those analytic contributions which are easiest to
determine. One can immediately rule out contributions from
operators of $\CO(p_\textrm{PGB}^4)$ because such operators
contain four derivatives, but tree-level matrix elements contain
only two fields upon which they can act.  One can also rule out
contributions from ${\cal O}(a^4)$ operators because, at
tree-level, they would produce terms without powers of masses in
them, and the chiral symmetry of the valence sector requires that
the kaon matrix element vanishes in the chiral limit.  Finally, all
contributions from $\CO(a^2 p_\textrm{PGB}^2)$ operators must be
proportional to $m^2_{xy}$ because the derivatives must act on the
two external kaons, giving a factor of $p^2$ which becomes
$m^2_{xy}$ when the kaons are on-shell.  This leads to the first
analytic term: $c_1 a^2 m_{xy}^2$.

All of the analytic terms that are not proportional to powers of
$a^2$ are the same as in the continuum partially quenched theory.
The mass dependence of these terms can be determined using CPS
symmetry \cite{Bernard:1985wf}, where C and P are the usual charge
conjugation and parity reversal symmetries, respectively. In QCD,
S corresponds to the exchange of $d$ and $s$ quarks, however, in
the mixed action lattice theory, we must impose a symmetry under
the interchange of $x$ and $y$ valence quarks instead: $x
\leftrightarrow y$, $m_x \leftrightarrow m_y$.  There are only two
linearly independent $\CO(m^2_q)$ terms allowed by this symmetry;
we choose to write them in the forms $c_2 m^4_{xy} \propto c_2
(m_x + m_y)^2$ and $c_3 (m_x-m_y)^2$. Operators of
$\CO(p_\textrm{PGB}^2 m_q)$ only contribute one new linear
combination of masses: $m^2_{xy} \textrm{Tr}(M_{\rm sea})$, where
$M_{\rm sea}$ is the $N_\textrm{sea} \times N_\textrm{sea}$  sea
quark mass matrix.

Finally, operators of $\CO(a^2 m_q)$ can be shown to give no new
independent contributions.  There are three possibilities for the
quark mass dependence: $(m_x + m_y)$, $(m_x - m_y)$, and
$\textrm{Tr}(M_{\rm sea})$.  The first term, $a^2(m_x + m_y)$, is
already included in $c_1$, the second, $a^2(m_x - m_y)$, is
forbidden by CPS symmetry, while the last term, $a^2
\textrm{Tr}(M_{\rm sea})$, vanishes for the following reason. The
factor $\textrm{Tr}(M_{\rm sea})$ always comes from the operator
$\textrm{Str}(\Sigma M^{\dag} + M \Sigma^{\dag})$ when $\Sigma=1$,
so the only operators that can lead to contributions of the form
$a^2\textrm{Tr}(M_{\rm sea})$ are $\CO(a^2)$ operators multiplying
$\textrm{Str}(\Sigma M^{\dag} + M \Sigma^{\dag})$. However, by
chiral symmetry, the $\CO(a^2)$ operators cannot generate
tree-level contributions to ${\cal M}_K$. Therefore, the
$a^2\textrm{Tr}(M_{\rm sea})$ terms also vanish.

To summarize, the following analytic terms contribute to \bk at
NLO:
\bea \left[ c_1 a^2m^2_{xy}, \quad \ c_2 m^4_{xy}, \quad \
c_3(m^2_X-m^2_Y)^2, \quad \ c_4 m^2_{xy}(m^2_{U_P} + m^2_{D_P} + m^2_{S_P})\right].
\eea Note that we have re-expressed them in terms of meson masses,
rather than quark masses, because those are what one measures in a
lattice simulation.

\subsection{Next-to-Leading Order \bk Results}
\label{sec:Results}

In this section we present results for a ``1+1+1" theory in which
$m_u \neq m_d \neq m_s$ in the sea sector, for a ``2+1" theory in
which $m_u = m_d \neq m_s$ in the sea sector, and for a ``full
QCD"-like expression in which we tune the valence-valence meson
masses equal to the taste-singlet sea-sea meson masses.

\bigskip

\bk at NLO in the 1+1+1 PQ theory is
\bea \label{eq:BK1+1+1}\left( \frac{B_K}{
B_0}\right)^{\textrm{PQ},1+1+1}&=&1+\frac{1}{16\pi^2 f_{xy}^2
m^2_{xy}}\left[I_{conn} +  I^{1+1+1}_{disc} \right] + c_1 a^2+ c_2
m^2_{xy} \nonumber \\ && + c_3\frac{(m^2_X-m^2_Y)^2}{m^2_{xy}}+c_4
(m^2_{U_P} + m^2_{D_P} + m^2_{S_P})\ .
\eea
The connected 1-loop contribution, $I_{conn}$, comes from
evaluating the integral in Eq.~\eqref{eq:Mconn}:
\bea  I_{conn}=
2m^4_{xy}\widetilde{\ell}(m^2_{xy})-\ell(m^2_X)(m^2_X+m^2_{xy})-
\ell(m^2_Y)(m^2_Y+m^2_{xy}), \eea
while the disconnected contribution, $I^{1+1+1}_{disc}$, comes from evaluating
the integral in Eq.~\eqref{eq:Mdisc}:
\bea
I^{1+1+1}_{disc}=\frac{1}{3}(m^2_X-m^2_Y)^2\frac{\partial}{\partial
m^2_X}\frac{\partial}{\partial m^2_Y}\left\{\sum_j
\ell(m^2_j)\left(m^2_{xy}+m^2_j
\right)R^{[4,3]}_j(\{M^{[4]}_{XY,I}\};\{\mu^{[3]}_I \}) \right\}.
\eea
In the above expressions, $\ell$ and $\tilde{\ell}$ are integrals regulated using
the standard \schpt\ scheme \cite{Aubin:2003mg,Aubin:2003uc}:
\bea \int \frac{d^4q}{(2\pi)^4}\frac{1}{q^2+m^2} \to
\frac{1}{16\pi^2}\ell(m^2), \\ \int
\frac{d^4q}{(2\pi)^4}\frac{1}{(q^2+m^2)^2} \to
\frac{1}{16\pi^2}\widetilde{\ell}(m^2), \eea
One can completely account for lattice finite volume effects by
turning the above integrals into sums.  This yields an additive
correction to the chiral logarithms \cite{Bernard:2001yj}:
\bea \ell(m^2)=m^2\left(\ln \frac{m^2}{\Lambda^2_{\chi}} +
\delta^{FV}_1(m\textrm{L})\right), \ \ \ \
\delta^{FV}_1(m\textrm{L})=\frac{4}{m\textrm{L}}\sum_{\vec{r} \neq
0} \frac{K_1(|\vec{r}|m\textrm{L})}{|\vec{r}|} \\
\widetilde{\ell}(m^2)=-\left(\ln \frac{m^2}{\Lambda^2_{\chi}}+1
\right) + \delta^{FV}_3(m\textrm{L}), \ \ \ \
\delta^{FV}_3(m\textrm{L})=2\sum_{\vec{r} \neq
0}K_0(|\vec{r}|m\textrm{L}) \eea
where the difference between the finite and infinite volume result
is given by $\delta_i^{FV}(m\textrm{L})$, and $K_0$ and $K_1$ are
modified Bessel functions of imaginary argument.

Finally, the residues and sets of meson masses that appear in the
1+1+1 disconnected contribution are defined to be:
\begin{eqnarray}\label{eq:Rdef} R^{[n,k]}_j(\{m\},\{\mu\})
&\equiv & \frac{\prod_{a=1}^k (\mu^2_a-m^2_j)}{\prod_{i\neq j}
(m^2_i-m^2_j)}, \\
\{M^{[4]}_{XY,I}\} &\equiv& \{m_X, m_Y, m_{\pi^0_I}, m_{\eta_I}
\}, \nonumber \\ \{\mu^{[3]}_I \} &\equiv& \{m_{U_I},m_{D_I},m_{S_I}
\}. \end{eqnarray}
and the mass eigenstates of the taste singlet flavor neutral PGB's
in the 1+1+1 case are\footnote{Strictly speaking, in the case
where $m_u\ne m_d$, the mass eigenstates of the flavor-neutral
sector are not the same as the physical states, $\pi^0_I$ and
$\eta_I$. Since the mixing between these two states is negligible
(and vanishes in the isospin limit), usually one does not make a
distinction between the mass eigenstates and the physical states.}
\bea m^2_{\pi^0_I,\eta_I} &=&
\frac{1}{3}\left[m^2_{U_I}+m^2_{D_I}+m^2_{S_I}\pm
\sqrt{m^4_{D_I}-(m^2_{U_I}+m^2_{S_I})m^2_{D_I}+m^4_{S_I} +
m^4_{U_I}-m^2_{S_I}m^2_{U_I} } \ \right]. \nonumber \\ && \eea

The expression in the 2+1 case is somewhat simpler:
\bea\label{eq:BK2+1} \left( \frac{B_K}{
B_0}\right)^{\textrm{PQ},2+1}&=&1+\frac{1}{16\pi^2 f_{xy}^2
m^2_{xy}}\left[I_{conn} +  I^{2+1}_{disc} \right]
+ c_1 a^2+ c_2
m^2_{xy}  \nonumber \\ &&  + c_3\frac{(m^2_X-m^2_Y)^2}{m^2_{xy}}+c_4
(2m^2_{D_P} + m^2_{S_P}) \ , \eea
where the connected term is the same as in the 1+1+1
case and the disconnected term is
\bea
I^{2+1}_{disc}=\frac{1}{3}(m^2_X-m^2_Y)^2\frac{\partial}{\partial
m^2_X}\frac{\partial}{\partial m^2_Y}\left\{\sum_j
\ell(m^2_j)\left(m^2_{xy}+m^2_j
\right)R^{[3,2]}_j(\{M^{[3]}_{XY,I}\};\{\mu^{[2]}_I \}) \right\},
\eea
\bea \{M^{[3]}_{XY,I}\} \equiv \{m_X, m_Y, m_{\eta_I} \},
\nonumber \\ \{\mu^{[2]}_I \} \equiv \{m_{D_I},m_{S_I} \}. \eea
When the up and down quark masses are degenerate, the flavor-neutral, taste-singlet mass
eigenstates are:
\begin{eqnarray} m^2_{\pi^0_I}  &= & m^2_{U_I} = m^2_{D_I}, \nonumber \\
m^2_{\eta_I} & = & \frac{m^2_{U_I}}{3} + \frac{2m^2_{S_I}}{3}.
\end{eqnarray}
\noindent The disconnected term also becomes simple enough that we choose to show
it explicitly:
\bea\label{eq:sea}
    I_{disc}^{2+1}=\frac{1}{3}
    \left(I_X+I_Y+I_\eta\right) ,
\eea
with
\bea I_X &&= \widetilde{\ell}(m^2_X)
\frac{(m^2_{xy}+m^2_X)(m^2_{D_I}-m^2_X)(m^2_{S_I}-m^2_X)}{(m^2_{\eta_I}-m^2_X)}
 \nonumber \\ && -\ell(m^2_X)\left[\frac{(m^2_{xy}+m^2_X)(m^2_{D_I}-m^2_X)(m^2_{S_I}-m^2_X)}{(m^2_{\eta_I}-m^2_X)^2}
 + \frac{2(m^2_{xy}+m^2_X)(m^2_{D_I}-m^2_X)(m^2_{S_I}-m^2_X)}{(m^2_Y-m^2_X)(m^2_{\eta_I}-m^2_X)} \right. \nonumber \\ &&
 \left. +\frac{(m^2_{D_I}-m^2_X)(m^2_{S_I}-m^2_X)-(m^2_{xy}+m^2_X)(m^2_{S_I}-m^2_X)
 -(m^2_{xy}+m^2_X)(m^2_{D_I}-m^2_X)}{(m^2_{\eta_I}-m^2_X)} \right], \eea
\bea I_Y=I_X(X \leftrightarrow Y),\eea
\bea I_\eta =
\ell(m^2_\eta)\frac{(m^2_X-m^2_Y)^2(m^2_{xy}+m^2_{\eta_I})(m^2_{D_I}-m^2_{\eta_I})(m^2_{S_I}-m^2_{\eta_I})}
{(m^2_X-m^2_{\eta_I})^2(m^2_Y-m^2_{\eta_I})^2}. \eea
Note that all of the sea quark dependence appears in the
disconnected terms, and that the sum of these terms vanishes for
degenerate valence quark masses.
It is clear that $I_\eta$ vanishes when $m_X = m_Y$, but it is not
immediately obvious that the other terms go to zero. However, it
can be shown that in the limit that $m_X \to m_Y$, $I_X \to -I_Y$
and thus the sum in Eq.~(\ref{eq:sea}) vanishes as claimed.

Multiple definitions of the ``full QCD" point appear in the mixed
action \chpt\ literature.  This is because the mixed
Ginsparg-Wilson valence, staggered sea theory has no true full QCD
point at finite lattice spacing.  Thus any choice of how to define
the full QCD point should be made for convenience. In this
paper, we consider the two cases that most closely resemble the
full unquenched theory.

One possible way to define full QCD for the mixed action theory
with 2+1 flavors is to set $m_X = m_{D_I} = m_{\pi^0_I}$ and $m_Y
= m_{S_I}$.  On the lattice, this is nontrivial because it
requires a tuning of the bare valence masses in order to set the
valence PGB masses to be those of the taste-singlet sea PGB
masses. This definition has the advantage, however, that the NLO
expression for \bk looks very much like the continuum expression,
except for the analytic term proportional to $a^2$. In this case,
the expression for $B_K$ at NLO reduces to
\bea
    \left( \frac{B_K}{ B_0}\right)^{\textrm{``full"}}
    &=& 1 + \frac{1}{16\pi^2 f_{xy}^2 m^2_{xy}}
    \Big[ 2m^4_{xy}\tilde\ell(m^2_{xy})
    +\frac{1}{2}(m^2_{X}-
    7m^2_{xy})\ell(m^2_{\eta_I})
    -\frac{1}{2} (m^2_{X}
    +m^2_{xy} )\ell(m^2_{X})  \Big]\nonumber\\
    &+& \tilde{c}_1 a^2+ c_2
    m^2_{xy}  +  c_3\frac{(m^2_X-m^2_Y)^2}{m^2_{xy}}
    +\tilde{c}_4 (2m^2_{X}+m^2_{Y})  ,
\eea where we have used the relationship $m_{\eta_I}^2 =
(4m^2_{xy} - m^2_{X})/3$ which holds in this limit.\footnote{Note
that the coefficients $\tilde{c}_1$ and $\tilde{c}_4$ are
different than those in Eqs.~(\ref{eq:BK1+1+1}) and
(\ref{eq:BK2+1}) because the mass-squared of the
taste-pseudoscalar meson and of the taste-singlet  meson differ by
a contribution of $\CO(a^2)$.} This clearly approaches the
standard result as $a\rightarrow 0$.

A popular, alternative way to define full QCD in MA\chpt\ is to
set the valence-valence meson mass equal to the pseudoscalar taste
sea-sea meson mass with the same quark content. For perfect G-W
valence quarks, this matching condition implies that the renormalized valence quark
mass equals the renormalized sea quark mass 
at tree level in $\chi$PT. This is 
the matching condition most often used in mixed action lattice simulations
since the Goldstone pion mass vanishes
in the chiral limit even at finite lattice spacing. Note that
although it is straightforward to set $m^{\rm val}_{ij} = m^{\rm
sea}_{ij}$ in the \chpt\ expressions, a lattice calculation using
domain wall fermions for the valence quarks would require a
non-trivial tuning, since the coefficient which renormalizes the
bare domain wall mass is different from that which renormalizes
the bare staggered quark mass. The tuning for this case has been
done on the MILC lattices by the LHP Collaboration
\cite{Renner:2004ck}, and they find that the renormalization
coefficients differ by around $30\%$. Our formula for $B_K$ to NLO
in MA$\chi$PT with this choice of tuning (to the taste
pseudoscalar) differs from the above tuning to the taste singlet
by terms of order $a^2$. Because no simplification occurs compared
to the most general formula,
Eq.~\eqref{eq:BK1+1+1}, we do not present a new expression for the
tuning to the taste pseudoscalar.

Given that the mixed action theory explicitly violates unitarity
at finite lattice spacing, there is no \emph{a priori} reason for
preferring one matching condition to another, and there are
numerous choices one could make. In the continuum limit, however,
all of the matching choices should be identical.  At fixed lattice
spacing, though, the two choices mentioned above have their
advantages and disadvantages.  The advantage of matching to the
taste singlet is that the theory is described by the full
continuum QCD formula (plus an $a^2$ analytic term), but the
disadvantage is that the taste singlet has the largest mass of the
16 taste pions, and on the MILC lattices this mass is still quite
large.  Matching to the taste pseudoscalar, the lightest of the
taste pions, then has the advantage of being closer to the
physical pion mass, but it has the disadvantage of having more
complicated $\chi$PT expressions.  However, once the explicit
$\chi$PT expressions exist, this is not much of a disadvantage. In
fact, given the complete partially quenched $\chi$PT expressions
it is advantageous to not use any tuning at all and to take
advantage of additional partially quenched data points in order to
best constrain the unknown parameters in the chiral fit.
This may not be true for all quantities, especially those for
which $\chi$PT is not a reasonable guide. In the following
numerical analysis we do not assume any matching condition has
been chosen; we analyze the results of our more general partially
quenched formula.

\section{Numerical Illustration of Mixed Action Contributions to \bk}
\label{sec:Num}

We plan on carrying out a mixed action numerical simulation of
$B_K$ using domain wall fermions on the publicly available MILC
improved staggered ensembles; we therefore use the known
parameters and previous measurements on these lattices in order to
obtain numerical error estimates for $B_K$ using our \chpt\ results.
Currently there are two lattice spacings with large statistics on
these ensembles, the ``coarse" MILC lattices with $a\approx 0.125$
fm and the ``fine" lattices with $a\approx 0.09$ fm.  In this
section we examine the modifications to the continuum expression
for $B_K$ that appear due to finite lattice spacing effects; these
include both taste-breaking errors from the staggered sea sector
and finite volume errors.

\bigskip

The NLO expression for \bk in a mixed action theory with $2+1$
flavors of sea quarks is given in Eq.~\eqref{eq:BK2+1}.
Discretization errors lead to two contributions -- the shift in
the mass-squared of the taste-singlet sea-sea meson that appears
in the 1-loop disconnected contribution and the analytic term
proportional to $a^2$.  Because we cannot \emph{a priori} know the
value of the coefficients of  the analytic terms, and because the
$\CO(a^2)$ analytic term does not arise purely from
taste-violating operators, we will neglect analytic terms in this
numerical analysis of the size of taste-breaking contributions in
$B_K$. We choose to study discretization errors on the $a\approx
0.125$ fm ``coarse" MILC lattices since taste violations will be
more pronounced than on the $a\approx 0.09$ fm lattices. In
particular, we use the parameters of the ensemble with the
lightest up and down sea quark masses on the smaller volume
($L/a=20$); this ensemble has a light quark mass of $am_l^{\rm
sea}=0.007$ and a strange quark mass of $am_s^{\rm sea} = 0.05$.

In order to estimate the size of discretization errors in $B_K$ we
calculate the percent difference between the 1-loop contributions
to \bk with and without taste-breaking:
\begin{equation}\label{eq:percdiff}
    \eta = \frac{B_K^{\rm 1-loop}(m_l^{\rm val}, a^2\Delta_I)-
    B_K^{\rm 1-loop}(m_l^{\rm val}, 0)}
    {B_K^{\rm 1-loop}(m_l^{\rm val}, 0)}\ .
\end{equation}
In this expression we have set the heavier valence bare quark mass
to be equal to the sea strange bare quark mass so that $\eta$ is a
function of the light valence quark mass, $m_l^{\rm val}$,  and
the taste-singlet splitting, $a^2\Delta_I$. The taste singlet
meson is the heaviest of all of the staggered sea-sea mesons, and
$a^2\Delta_I$ is approximately $(450~\text{MeV})^2$ on the coarse
lattices \cite{Aubin:2004wf}.  Because the only sea-sea mesons
that contribute to the \bk at 1-loop in the mixed action theory
are taste-singlets, this large splitting makes the effective sea
quark mass considerably larger than a nominal light sea quark mass
of $m_s/10$ or $m_s/7$ would suggest. On the fine lattices this
splitting is less, close to
$(280~\text{MeV})^2$, which scales appropriately according to
$a^2\alpha_s^2$ \cite{Aubin:2004wf}, and this shows that it is
necessary to approach the continuum limit in order to approach the
chiral limit in the sea sector.  Note, however, that the sea quark
dependence is predicted to be small in the non-analytic
contribution to our formulas.  The sea quarks only contribute to
the disconnected hairpin diagrams in $B_K$, and this is only
around $15\%$ of the connected piece at the physical point.  It
will, however, be necessary to study the numerical data in order
to determine the size of the analytic contribution, as well as to
test the validity of our \chpt\ formulas at the physical strange
quark mass.

\begin{figure}[t]
\begin{center}
\includegraphics[width=5in]{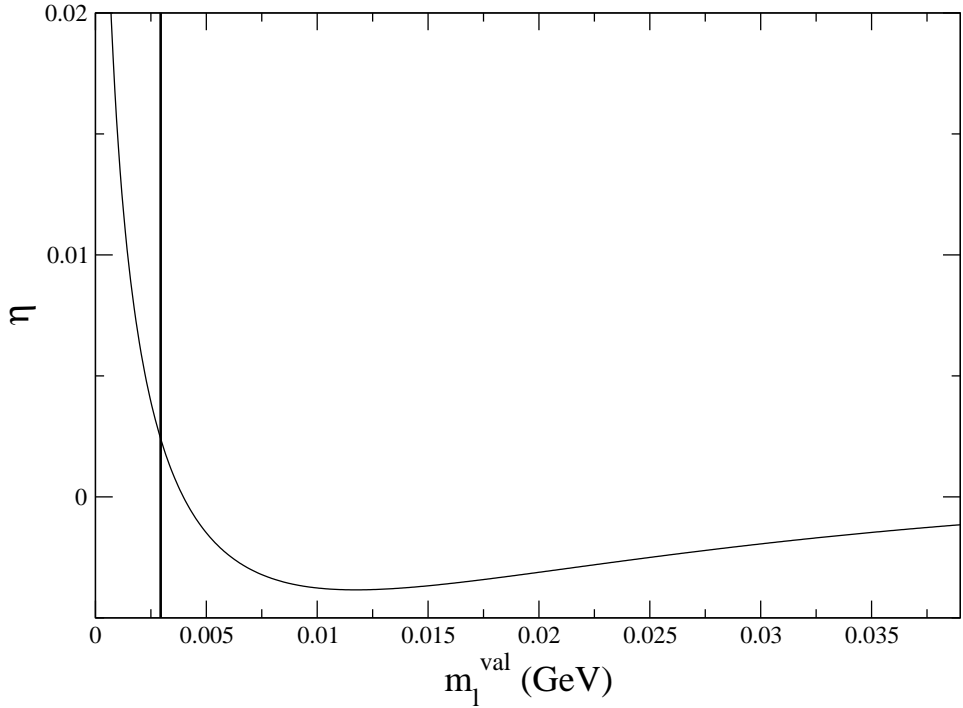}
\caption{Percent difference between the 1-loop contributions to
\bk with and without taste-breaking discretization errors,
Eq.~\eqref{eq:percdiff}, as a function of valence light quark
mass. The masses and taste splittings are those of the MILC coarse
ensemble ($a=0.125$ fm) with $am_{l}=0.007$ and $am_{s}=0.05$. The
vertical line shows the physical light quark mass.}
\label{fig:ContDiffplot}
\end{center}
\end{figure}

We plot the percent difference, Eq.~(\ref{eq:percdiff}), as a function of valence quark mass in
Fig.~\ref{fig:ContDiffplot}, setting $a^2\Delta_I$ to be the value measured in MILC
simulations on the coarse lattices \cite{Aubin:2004wf}.
In this plot, the vertical line shows the location of the physical
value of the average up/down quark mass, $m_l^{\rm phys}\approx
m_s/27$. One can see that for larger valence light masses, $\eta$
rapidly goes to zero. This is to be expected as the difference
between the mass-squared of purely valence and purely sea mesons
will ultimately be negligible for sufficiently large quark masses.
At quark masses below $m_l^\textrm{val}\approx 0.01$ GeV, $\eta$
begins to increase as the valence light mass decreases, such that
as $m_l^{\rm val}\to 0$, the percent difference blows up rapidly.
Note, however, that this dramatic increase does not happen until
below the physical mass, so in the region of interest
($m_l^{\rm val}\ge 0.002$ GeV) the error coming from taste
violations is never higher than $0.5\%$. Note also that this
estimate depends upon the value of the cutoff, $\Lambda_\chi$,
used (Fig.~\ref{fig:ContDiffplot} uses $\Lambda_\chi=1$GeV),
although any cutoff dependence can be absorbed into the analytic
terms which are not included in this numerical analysis.  Varying
$\Lambda_\chi$ within the range of $0.5$ GeV to $1.5$ GeV changes
this picture in the relevant light quark mass range only by a
nearly constant vertical shift at the half a percent level.

As for the analytic terms, if all terms up to NLO were included in
Eq.~\eqref{eq:percdiff} this ratio would be explicitly independent
of those analytic terms present in the continuum. The remaining
term which is proportional to $a^2$ has been set to zero since we
do not know \emph{a priori} its value. In our analysis, however,
by choosing a non-zero value for $c_1$, our plot in
Fig.~\ref{fig:ContDiffplot} would just shift vertically by a
nearly constant amount as a function of the quark mass in the
region of interest.  If the scaling dependence of quenched domain
wall fermions with various gauge actions is any guide (see Fig. 7
of Ref.~\cite{Lee:2006cm}), this term will give a contribution
that is on the order of a few percent.

We now repeat the above analysis, but include errors due to the
finite spatial extent of the lattice. Such finite volume effects
can be quite noticeable at the lightest sea quark masses available
on the MILC configurations. One might imagine that the finite
volume effects in the mixed case would not be very different than
the continuum case, since the taste violating effects only enter
through the taste singlet meson, which has a large mass. In
actuality, this is precisely where there could be a problem, since
the heavy singlet mass appears only in the sea sector.  Partially
quenched pathologies begin to appear when $m^{\rm val}_\pi <
m^{\rm sea}_\pi$.\footnote{See, for example,
Ref.~\cite{Bernard:1993sv}.} Consequently, if the sea mesons are
significantly heavier than the valence mesons (as they are in the
mixed action theory) these pathologies may become more pronounced.
We will see that this is, in fact, the case.

\begin{figure}[t]
\begin{center}
\includegraphics[width=5in]{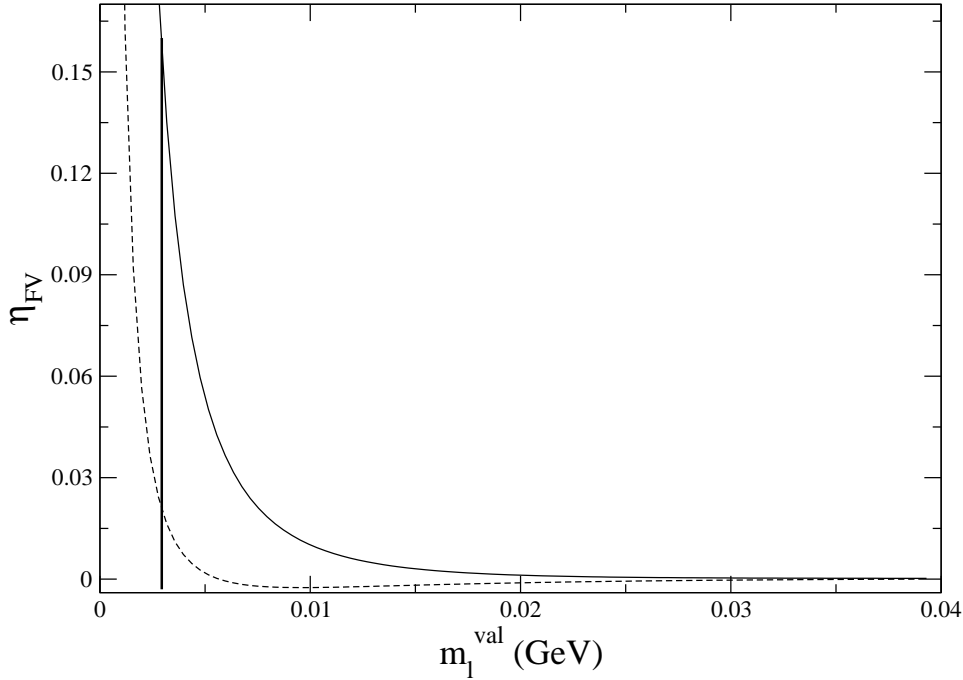}
\caption{Percent difference in the 1-loop contributions to \bk in
finite and infinite volume, Eq.~\eqref{eq:FVratio}, as a function
of valence light quark mass. The dashed curve corresponds to
$\eta_{FV}$ with $a^2\Delta_I=0$ (the continuum case), and the
solid curve shows $\eta_{FV}$ with $a^2\Delta_I$ set to its value
on the coarse ($a=0.125$ fm) MILC lattices. The sea quark masses
are $am_{l}=0.007$ and $am_{s}=0.05$, and the spatial extent of
the lattice is $L=20$.} \label{fig:FVplot}
\end{center}
\end{figure}

In analogy with Eq.~\eqref{eq:percdiff}, we define $\eta_{FV}$ to
be the percent difference between the 1-loop contribution to \bk
in the mixed theory at finite volume and \bk in the mixed theory
at infinite volume, both including discretization errors:
\begin{equation}\label{eq:FVratio}
    \eta_{FV}(m_l^{\rm val}, a^2\Delta_I)
    = \frac{B_K^{\rm 1-loop,FV}(m_l^{\rm val}, a^2\Delta_I, L)-
    B_K^{\rm 1-loop}(m_l^{\rm val}, a^2\Delta_I)}
    {B_K^{\rm 1-loop}(m_l^{\rm val}, a^2\Delta_I)}\ .
\end{equation}
We evaluate the above expression at a spatial lattice size of
$L=20$; the remaining parameters are the same as in the previous
analysis. We then plot in Fig.~\ref{fig:FVplot} two curves -- the
dashed curve shows the percent difference in
Eq.~\eqref{eq:FVratio} for the continuum limit,
$\eta_{FV}(m_l^{\rm val}, 0)$, while the solid curve shows the
same percent difference with $a^2\Delta_I$ set to its value on the
coarse MILC lattices. Again, the vertical line indicates the
physical light quark mass.

One can see that, for the region
$m_l^{\rm val} \ge 0.01$ GeV, the error associated with finite
volume effects, while not negligible, is quite small, of
$\CO(1\%)$ or less. Although the continuum and finite lattice
spacing cases are different, in each instance finite volume
effects are small.  As the valence light quark mass drops below
the sea light quark mass, the error shoots up rapidly, which is
expected because quenching artifacts (where finite volume effects
are more pronounced) become more noticeable in this region. As
discussed above, the mixed case sees these quenching artifacts at
a larger valence mass, since the sea mesons are heavier.  Although
one might worry about this significant difference between the
continuum and the mixed cases for the lighter masses (the
difference is rather striking at the physical mass), this is not a
practical problem.  For example, the MILC ensemble with a light
sea quark mass of $am_l^{\rm sea} = 0.005$ has a spatial length of
$L=24$ as opposed to 20 for the heavier sea quark mass ensembles.
This increase in volume for the lighter quark mass was chosen by
MILC to reduce finite volume effects for such a light sea quark
mass \cite{Aubin:2004fs}.
In our planned simulations of $B_K$, the quantity which we wish to
keep large is the combination $m^{val}_\pi L$, which as a general
rule-of-thumb should be 4 or more to keep the finite volume
effects relatively small.\footnote{Although finite volume $\chi$PT at can be used to correctly describe more significant finite size errors, if the removable 1-loop finite volume corrections to \bk are around 15\%, then it is likely that the remaining 2-loop finite volume errors will still be a few percent.  Such a large systematic uncertainty from finite size effects is unacceptable if one is aiming for an overall error of 5\% in $B_K$.} Note though, that in
Fig.~\ref{fig:FVplot}, the renormalized valence quark mass
$m_l^\textrm{val} = 0.01$ GeV, and with L=20 this corresponds to
an $m_\pi L\approx 3$, leading to a finite volume relative error
of about $1.5\%$.  As the valence quark mass (and thus the valence
pion mass) decreases, this error goes up rapidly. If we want to go
to lighter quark masses, we will need to use the MILC lattices
with larger volume (L=24) so that $m_\pi^{val} L$ does not become
significantly smaller than 3, and the finite volume corrections
stay below the $2\%$ level. The key point is that simulations are
done to the right of this ``wall'' in Fig.~\ref{fig:FVplot} at
which the error explodes, so one can correct for finite volume
errors before performing extrapolations to the continuum and
physical light quark mass.

\section{Conclusions}
\label{sec:Conc}

In this work we have calculated the expression for $B_K$ in a
mixed action lattice theory with Ginsparg-Wilson valence quarks
and staggered sea quarks to next-to-leading order in chiral
perturbation theory.  We have discussed in some detail how to
extend the continuum calculation to the mixed action case, and we
have provided expressions for both a ``1+1+1" partially quenched
theory ($m_u \neq m_d \neq m_s$) and a ``2+1" partially quenched
theory ($m_u = m_d \neq m_s$), both of which reduce to the
corresponding partially quenched QCD expressions in the continuum
limit.

It is illustrative to compare our expression for \bk in mixed
action chiral perturbation theory to that for other lattice
theories.  Four parameters are needed to describe $B_K$ in the
continuum: one leading order constant, $B_0$, and three NLO
coefficients.   In the case of pure Ginsparg-Wilson lattice
fermions, the expression for \bk contains one additional
coefficient proportional to $a^2$.  For domain-wall lattice
fermions there is an additional constant, $m_\textrm{res}$, which
comes from chiral symmetry breaking due to the finite domain wall
separation. This term simply enters \bk as an additive shift to
the quark mass and can be separately measured in a tree-level fit
to the pion mass-squared.  In the mixed action lattice theory with
staggered sea quarks and domain wall valence quarks that we have
considered here, taste-symmetry breaking effects produce an
additive shift to the sea-sea meson mass squared.  This is the
only new term that appears in the calculation of $B_K$ with
domain-wall quarks on a staggered sea as compared to a pure
domain-wall calculation.  In contrast, the expression for \bk with
staggered valence quarks on a staggered sea contains many new
parameters, each of which must be determined from lattice
simulations and subsequently removed in order to extract the value
of \bk in the continuum. It is interesting to note that the
expression for \bk in the mixed G-W, staggered lattice theory is
no more complicated  than that for a domain-wall simulation in
which a different value of the domain-wall separation is used in
the valence and sea sectors;  such a ``mixed" domain-wall
simulation was previously proposed in
Ref.~\cite{Golterman:2004mf}. In this case the value of
$m_\textrm{res}$ would differ in the valence and sea sectors, and
the corresponding expression for \bk could be gotten from our
expression, Eq.~(\ref{eq:BK1+1+1}), by simply making the
replacements $m_\textrm{res} \to m_\textrm{res}^\textrm{valence}$
and  $a^2 \Delta_I \to m_\textrm{res}^\textrm{sea}$.  Thus the
taste-singlet sea-sea meson mass shift can be thought of as an
effective $m_\textrm{res}$ in the sea sector, though this
effective ``$m_\textrm{res}$" scales as $a^2$, and consequently
vanishes in the continuum limit.

Finally, we have presented a numerical analysis of the resulting
expressions in which we have examined the size of discretization
errors from taste-symmetry breaking in the sea sector and finite
volume errors for the MILC coarse ($a\approx 0.125$ fm, L=20 and
24) lattice ensembles. We find that the non-analytic
taste-breaking contributions to \bk in the mixed action theory are
around $0.5\%$ over the range of the extrapolation and so are
quite small. The finite volume effects are somewhat larger for the
mixed action case than in the continuum, but still remain at or
below the $2\%$- level for the values of the light quark masses
used to generate the MILC ensembles. It will of course be
necessary to study the numerical lattice data in order to
determine the size of the analytic contribution, as well as to
test the validity of our NLO \chpt\ formulas at the physical
strange quark mass.

A lattice calculation of $B_K$ using domain-wall
valence quarks on top of improved staggered field configurations
combines the best properties of both fermion discretizations.
This method will be competitive with other established
methods for calculating $B_K$, and ultimately it should give a
useful constraint on the CKM matrix and phenomenology.

\section*{Acknowledgments}
We thank Andreas Kronfeld, Donal O'Connell and Andr\'e Walker-Loud
for useful discussions and careful readings of the manuscript. We are grateful to Steve Sharpe and Claude Bernard for useful comments. We
also thank the Institute for Nuclear Theory at the University of
Washington for their hospitality while some of this work was
completed. CA would like to thank Norman Christ for useful discussions. This research was supported by the DOE under grant nos.
DE-FG02-92ER40699 and DE-AC02-76CH03000.


\bibliography{SuperBib}

\end{document}